\documentclass[numberedappendix]{emulateapj}
\usepackage{apjfonts}
\usepackage{amsmath}
\usepackage{multirow}
\usepackage{array}
\usepackage[usenames]{color}
\allowdisplaybreaks[1]
\usepackage[utf8x]{inputenc}

\newcommand{\comment}[1]{}
\definecolor{purple}{RGB}{160,32,240}

\newcommand{\Msun}{M_{\odot}}

\newcommand{\Reff}{R_\mathrm{e}}

\newcommand{\rvir}{R_\mathrm{vir}}

\newcommand{\plotgrace}[1]{\includegraphics[width=\columnwidth,type=pdf,ext=.pdf,read=.pdf]{#1}}

\newcommand{\plotminigrace}[1]{\includegraphics[width=0.5\columnwidth,type=pdf,ext=.pdf,read=.pdf]{#1}}

\begin{document}

\shortauthors{BEHROOZI, RAMIREZ-RUIZ, FRYER}
\shorttitle{Interpreting Short Gamma-Ray Bursts}
\title{Interpreting Short Gamma Ray Burst Progenitor Kicks and Time Delays Using the Host Galaxy---Dark Matter Halo Connection}
\author{Peter S. Behroozi}
\affil{Space Telescope Science Institute, Baltimore, MD 21218, USA}

\author{Enrico Ramirez-Ruiz}
\affil{Department of Astronomy and Astrophysics, University of California Santa Cruz, CA 95064, USA}

\author{Christopher L.\ Fryer}
\affil{Los Alamos National Laboratory, Los Alamos NM, 87545, USA}

\begin{abstract}
Nearly $20\%$ of short gamma-ray bursts (sGRBs) have no observed host galaxies.  Combining this finding with constraints on galaxies' dark matter halo potential wells gives strong limits on the natal kick velocity distribution for sGRB progenitors.  For the best-fitting velocity distribution, one in five sGRB progenitors receives a natal kick above 150 km s$^{-1}$, consistent with merging neutron star models but not with merging white dwarf binary models.  This progenitor model constraint is robust to a wide variety of systematic uncertainties, including the sGRB progenitor time-delay model, the \textit{Swift} redshift sensitivity, and the shape of the natal kick velocity distribution.  We also use constraints on the galaxy---halo connection to determine the host halo and host galaxy demographics for sGRBs, which match extremely well with available data.  Most sGRBs are expected to occur in halos near $10^{12}\Msun$ and in galaxies near $5\times10^{10}\Msun$ ($L_\ast$); unobserved faint and high-redshift host galaxies contribute a small minority of the observed hostless sGRB fraction.  We find that sGRB redshift distributions and host galaxy stellar masses weakly constrain the progenitor time-delay model; the active vs.\ passive fraction of sGRB host galaxies may offer a stronger constraint.  Finally, we discuss how searches for gravitational wave optical counterparts in the local Universe can reduce followup times using these findings.
\end{abstract}


\section{Introduction}

\label{intro}

Gamma-ray bursts \citep{Klebesadel73} show a strong bimodality in their prompt duration distributions \citep{Kouveliotou93}.  Long gamma-ray bursts, with $T_{90}$ (the time for 90\% of their photons to be received) greater than 2s, have been associated with Type Ic supernovae (\citealt{Galama98,Bloom02b,Berger11}; see also references in \citealt{Levesque13}).  For short gamma-ray bursts (sGRBs; $T_{90} < 2$s), several progenitor classes have been proposed.  These include mergers of compact stellar remnants (i.e., combinations of neutron stars and black holes) via gravitational radiation, accretion-induced collapse of neutron stars, magnetar flaring, and magnetar formation through binary white dwarf mergers or white dwarf accretion-induced collapse \citep[see][and references therein]{Nakar07,Lee07,Gehrels09,Berger11b,Faber12,Gehrels13,Berger13}.

These progenitor class models have many distinguishing signatures.  The redshift distribution of sGRBs strongly depends on the delay time between star formation and gamma-ray emission \citep{Guetta06,Hao13}.  For models with longer delay times (e.g., binary mergers), the peak of the sGRB redshift distribution will occur much later than the peak in the cosmic star formation rate; the opposite is true for prompt sGRB channels (e.g., magnetar collapse).  Longer delay times also imply that the host galaxy masses will be higher (as the host galaxies have had more time to grow; \citealt{Leibler10}), the specific star formation rates will be lower \citep{Zheng07GRB,Berger09}, and the morphology distribution will be more elliptical (as more massive and redder galaxies are more elliptical; \citealt{Lintott08}); see \cite{Berger11} for a review.

Another important observational signature is the \textit{offset} between the sGRB location and the host galaxy where the sGRB progenitors were formed.  The conversion of stars to black holes, neutron stars, or white dwarves imparts a significant velocity kick to the stellar remnant, which can be on the order of hundreds of kilometers per second \citep{Hansen97,Fryer99}.  If the remnant is in a binary system that survives the velocity kick, the binary system may well be expelled from its host galaxy \citep{Zheng07GRB,Zemp07}.\footnote{E.g., the escape velocity for two 1.4 solar-mass neutron stars a solar radius apart is 731 km s$^{-1}$.  By comparison, the velocity required to escape from the center of a $10^{12}\Msun$ dark matter halo (assuming an NFW potential \citealt{NFW97} with concentration $c=10$) to its virial radius \citep{mvir_conv} is only 444 km s$^{-1}$.}  For long gamma-ray bursts, the \textit{lack} of significant differences between the burst locations and the host galaxies' stellar distributions provided an early indication that binary mergers were not long gamma-ray burst progenitors \citep{Bloom02a}.  On the other hand, \textit{Swift} and \textit{Hubble Space Telescope} follow-up observations have found large offsets between sGRBs and nearby galaxies in a significant fraction of cases \citep{Prochaska06,Fong10,Berger10,Fong13,Fong13b}.  E.g., \cite{Fong13b} find that $\approx 20\%$ of sGRBs occur at more than 5 effective radii from the nearest likely host galaxy.

These observations would seem to strongly favor binary mergers as sGRB progenitors \citep{Lee05,Church11,Fong13b,Berger13}, yet it is not straightforward to interpret the observed offsets.  For example, sGRB host galaxies with low luminosities and/or high redshifts may be below the detection threshold for observation \citep{Berger10}, meaning that the inferred offsets to the nearest observed galaxy would be much too high.  Because it is often impossible to directly measure the depth of the potential well surrounding a galaxy, it is also difficult to robustly estimate the velocity distribution of the sGRB progenitors from the observed offsets \citep{Bloom06}.  Finally, mergers which statistically trace the globular cluster distribution around galaxies \citep{Grindlay06,Lee10,Church11,Samsing13} would result in very different interpretations for the offsets than if mergers traced the galactic potential \citep{Fryer99,Bloom99,Rosswog03,Belczynski06}.

Using more detailed modeling to confirm the current interpretation of the offsets as binary neutron star mergers or neutron star--black hole mergers \citep{Fong13b} is therefore crucial, especially as this interpretation implies that optical counterparts exist for an important class of gravitational wave sources \citep{Lattimer74,Rosswog99,Metzger10,Roberts11,Bauswein13,Metzger13,Berger13,Kasen13,Barnes13,Tanaka13,Grossman13}.\footnote{Detection of r-process powered transients might provide additional important evidence in support of the merger hypothesis; \cite{Tanvir13} report tentative evidence for such a signature in GRB130603B.}

In this paper, we present a framework for modeling sGRBs which can predict the host galaxy luminosities, the host galaxy offsets, and the host dark matter halo masses for a wide range of assumptions about the progenitor velocity kick and time delay distributions.  This framework takes advantage of recent empirical constraints on the galaxy stellar mass --- halo mass relation as well as on galaxy star formation histories as a function of halo mass and redshift \citep{BWC13}.  Galaxy star formation histories convolved with the assumed time delay distribution then give sGRB rates as a function of halo mass and redshift; as the dark matter halo mass determines the gravitational potential well, the sGRB offset distribution can be straightforwardly calculated from the assumed velocity kick distribution.  In addition, the sGRB host galaxy luminosity distribution can be calculated by convolving galaxy star formation histories with a stellar population synthesis model \citep{conroy-09,Conroy10}.  All of these results are provided in an observational context; i.e., including observational systematic effects on the redshift distribution and observed host properties of the modeled sGRBs.

We describe the constraints on host halo masses for short GRB progenitors in \S \ref{s:halo_masses}, and show examples for several different assumed sGRB delay distributions.  We place upper limits on sGRB---galaxy offsets as well as make comparisons with current observations in \S \ref{s:hostless}.  We discuss implications in \S \ref{s:discussion} and summarize our conclusions in \S \ref{s:conclusions}.  In this work, we assume a flat, $\Lambda$CDM cosmology with parameters $\Omega_M = 0.27$, $\Omega_\Lambda = 0.73$, $h=0.7$, $n_s = 0.95$, and $\sigma_8 = 0.82$.

\begin{figure}
\plotgrace{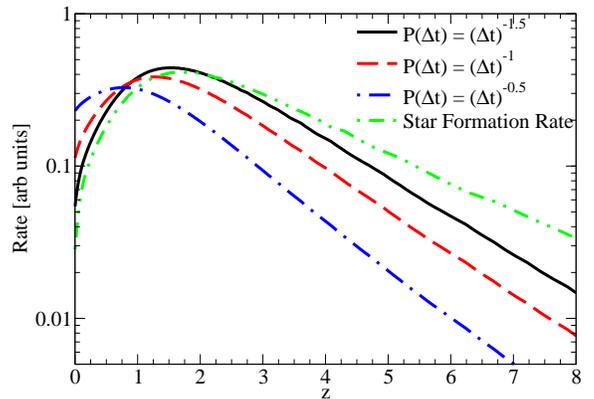}
\caption{The cosmic sGRB rate (i.e., the rate of sGRBs per unit comoving volume) as a function of redshift for the three time-delay distributions studied in this paper.  For these models,  the sGRB rate at $z\approx 1.5$ compared to the sGRB rate at $z\approx 0$ represents the clearest observable difference.  Note that this figure depends only on collected observations of the cosmic star formation rate \citep[green line; see][]{BWC13} and does not depend on the halo model assumed.}
\label{f:grb_rate}
\end{figure}

\begin{figure*}
\plotgrace{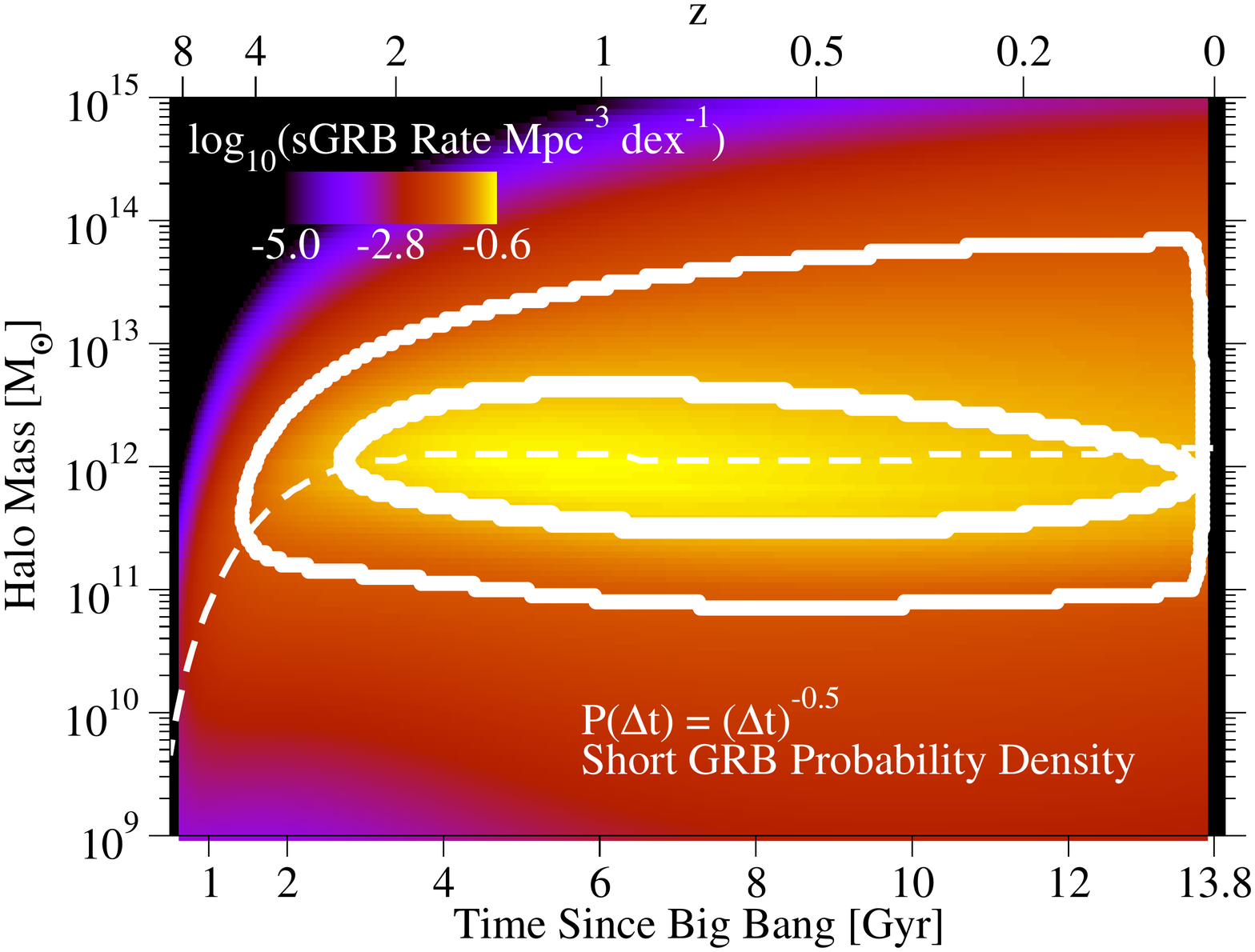}\plotgrace{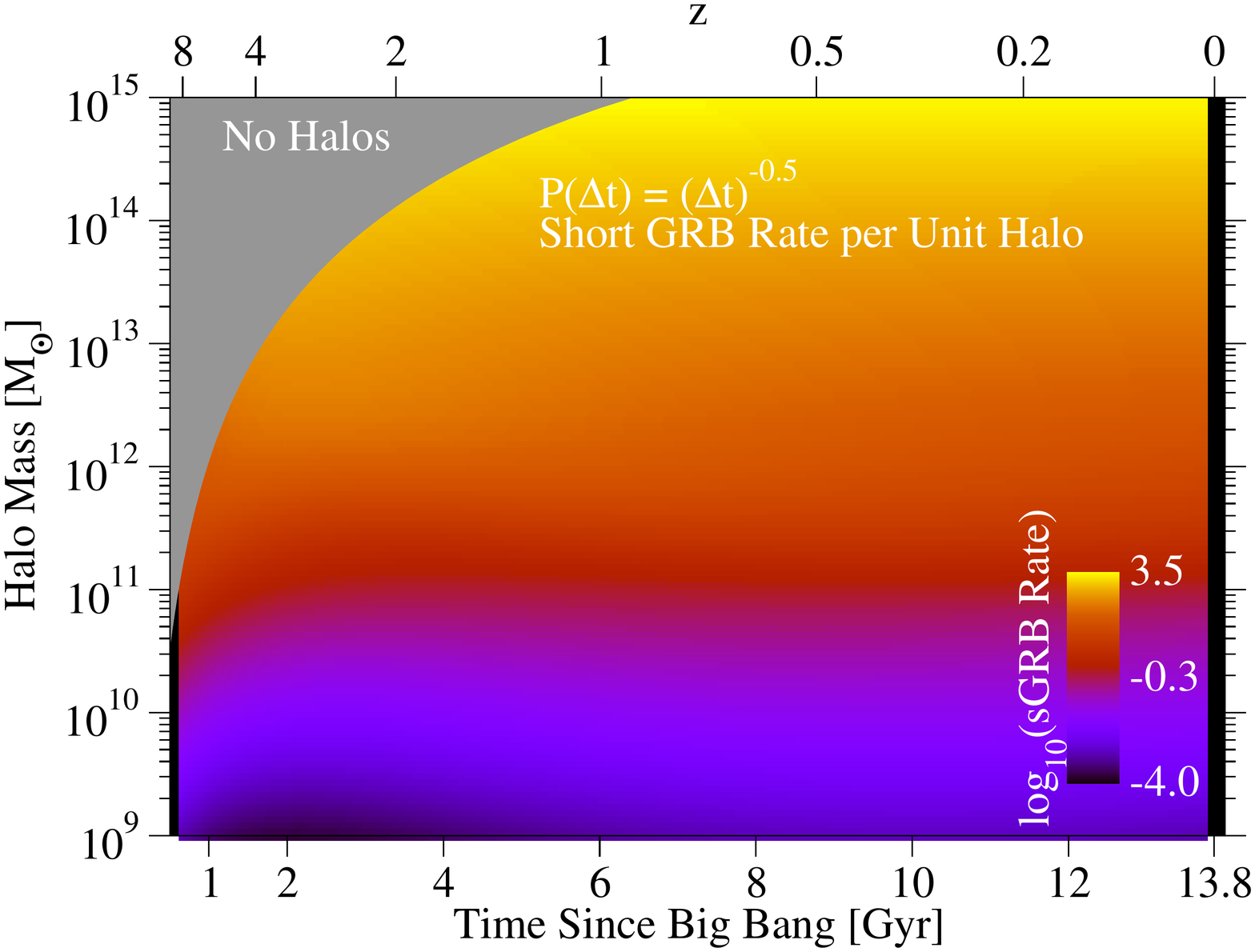}
\plotgrace{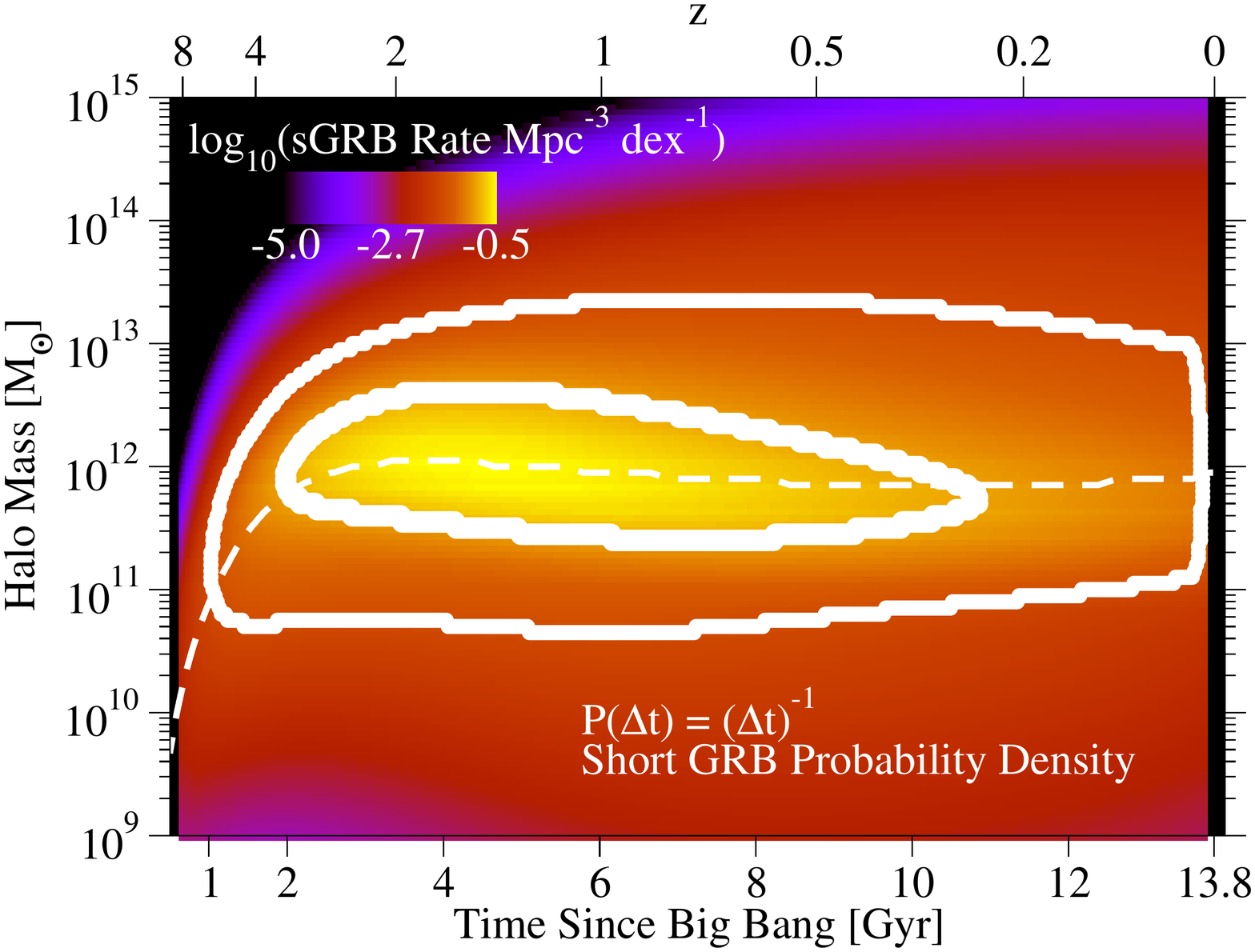}\plotgrace{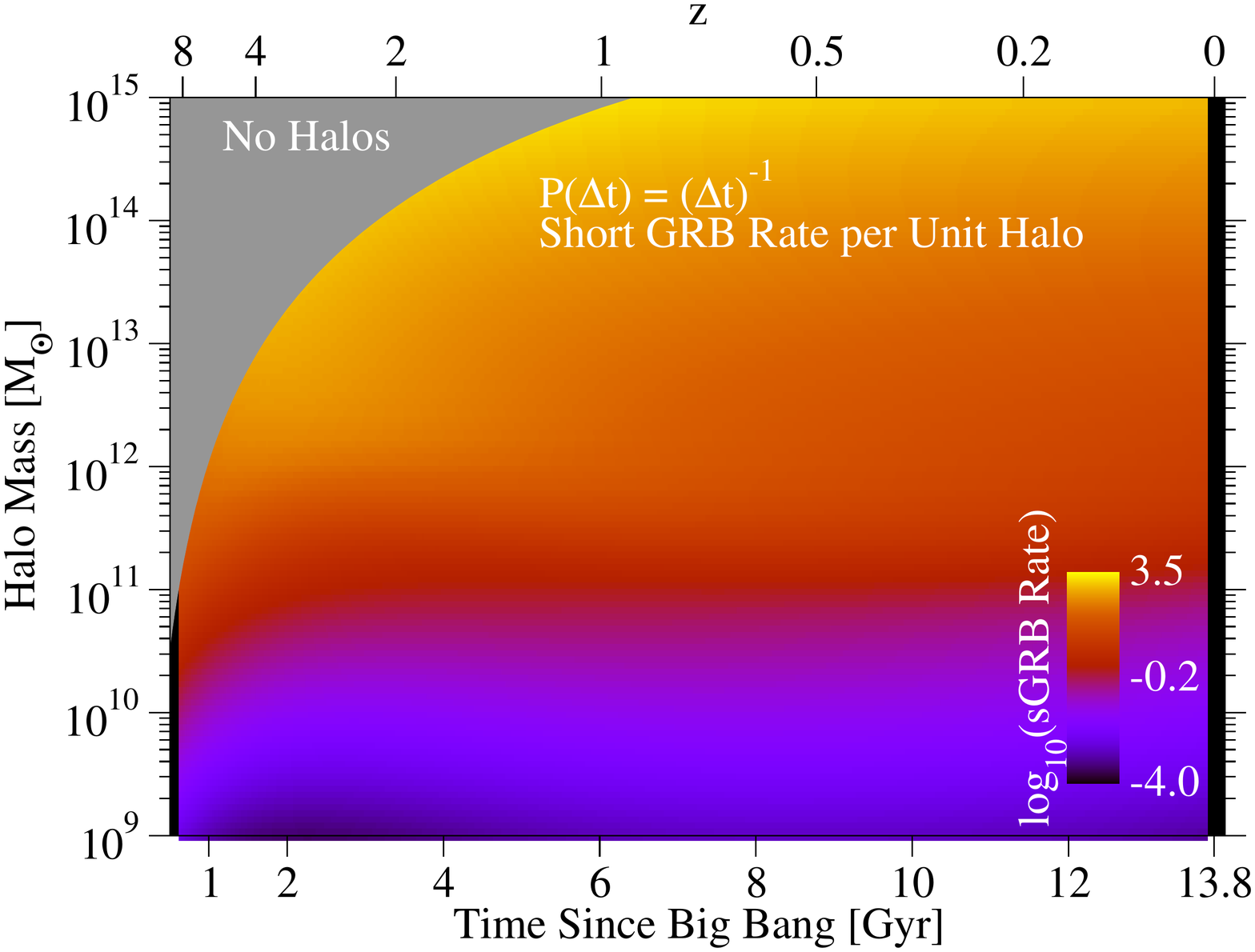}
\plotgrace{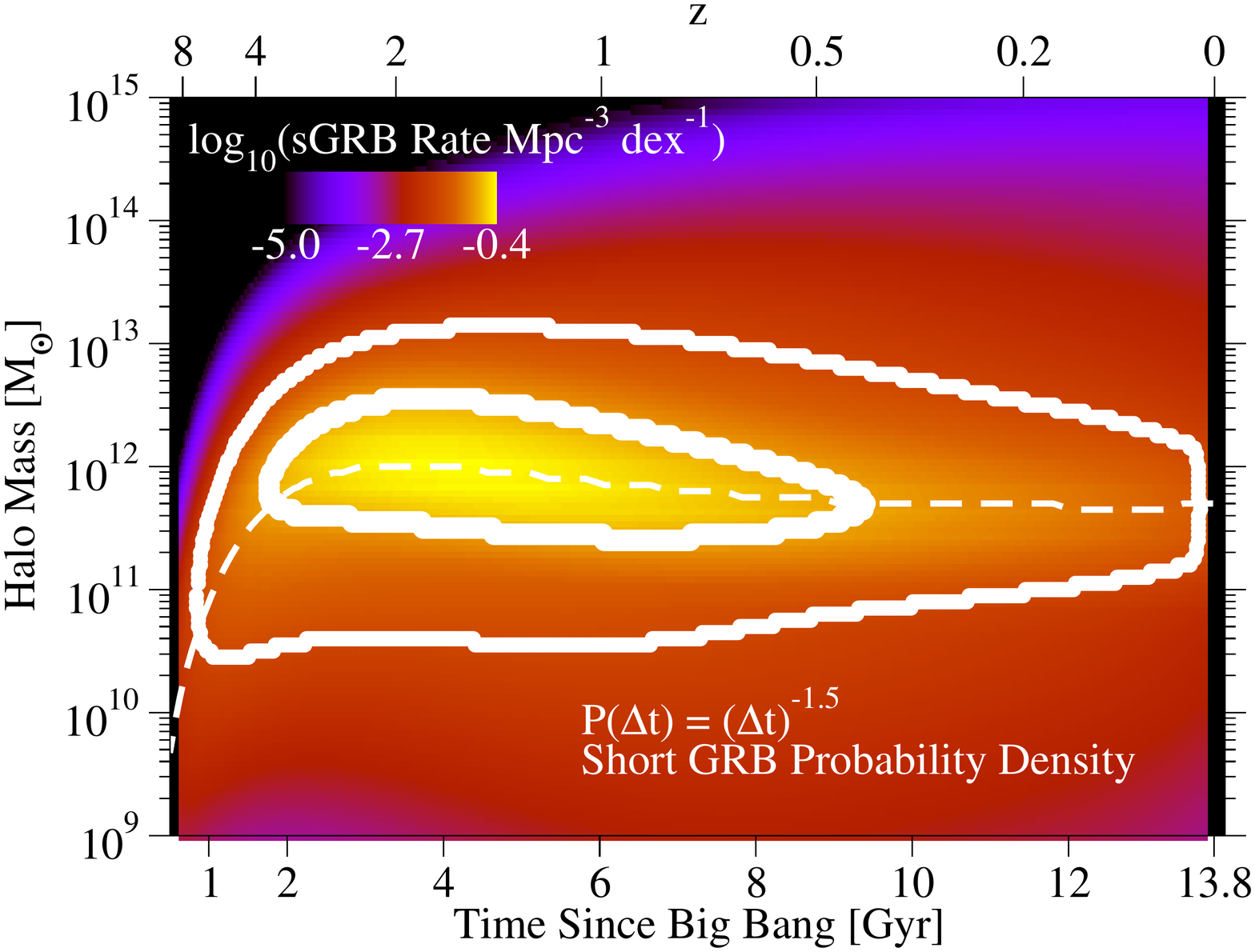}\plotgrace{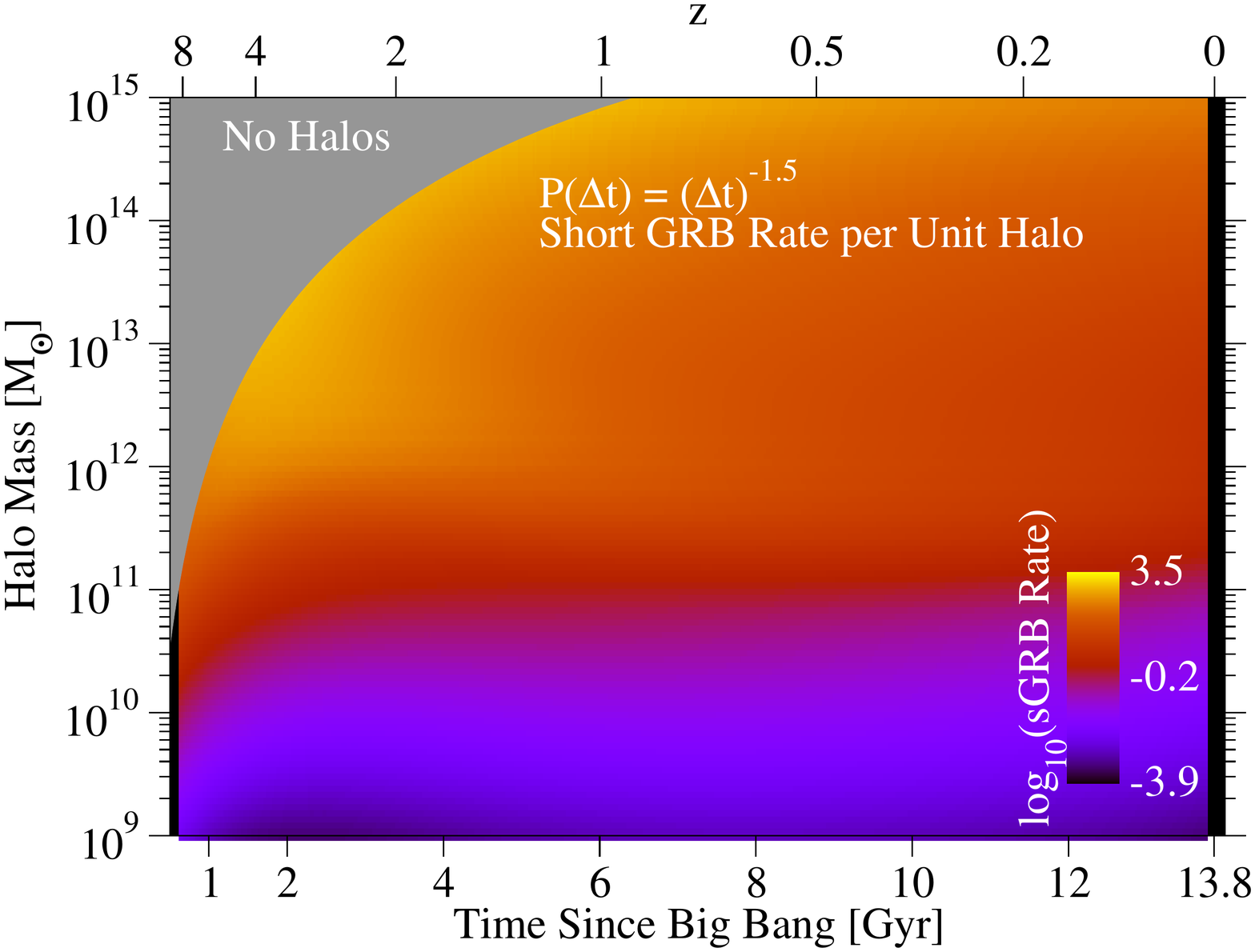}
\caption{Host halo statistics for sGRBs as a function of redshift and halo mass (\S \ref{s:halo_methods}). \textbf{Left panels} show joint probability distributions for sGRBs as a function of host halo mass and redshift; the units are sGRBs per unit time per unit comoving volume per dex in halo mass.  Thick and thin white contours enclose 50\% and 90\% of sGRBs, respectively.  The white dashed line shows the median halo mass for sGRBs as a function of redshift.  The \textbf{top}, \textbf{middle}, and \textbf{bottom} panels show the distributions for time-delay power-law indices of $-0.5$, $-1.0$, and $-1.5$, respectively.  \textbf{Right panels} show the individual sGRB rate per unit halo as a function of halo mass and cosmic time; the units are sGRBs per unit time.  Arrangement from top to bottom is analogous to the left-hand panels.  Even though the sGRB rate per unit halo is very high for massive halos, the rarity of massive halos means that most sGRBs appear in $\approx 10^{12}\Msun$ halos (see left-hand plots).  The normalization of all panels is arbitrary, due to the unknown beaming angles for sGRBs.}
\label{f:grb_halos}
\end{figure*}

\section{Host Halo Masses}
\label{s:halo_masses}

\subsection{Methods}

\label{s:halo_methods}

Short GRBs are expected to trace the star formation rate of galaxies, albeit with a time delay between star formation and the gamma-ray burst which depends on the progenitor class \citep[e.g.,][]{Lee07,Gehrels09,Berger13}.  Both the stellar mass and star formation rate of galaxies have been shown to be tightly linked to the galaxy's host dark matter halo mass and the cosmological redshift \citep[e.g.,][]{cw-08,Leitner11,Wang12,Moster12,Bethermin13,BWC13,Yang13,Lu13}.  To calculate the expected short GRB rate per dark matter halo, we therefore adopt the best-fitting mean galaxy star formation rate as a function of halo mass and redshift, $SFR(M_h, z)$, from \cite{BWC13}, and consider several options for the delay time distribution.  For reference, the derivation of $SFR(M_h, z)$ is briefly discussed in Appendix \ref{s:sfr}.

A growing body of evidence suggests that the sGRB delay-time probability distribution, $P(\Delta t)$, has a power-law form with $P(\Delta t) \propto (\Delta t)^{-1}$; however, some uncertainty remains on the exact exponent \citep[and references therein]{Berger13}.  A $(\Delta t)^{-1}$ distribution formally diverges as $\Delta t$ approaches 0, so we consider time-delay distributions which are power laws with an initial time cutoff:
\begin{equation}
\label{e:prob}
P(\Delta t) \propto \left\{ \begin{matrix} (\Delta t)^n & \textrm{if }\Delta t > t_\mathrm{cut}\\ 0 & \textrm{otherwise} \end{matrix} \right.
\end{equation}
The choice of $t_\mathrm{cut}$ has little effect for reasonable values.  For example, with a power law index of $n=-1$, the mean time delay for GRBs which happen within the Hubble time $t_H$ is
\begin{equation}
\langle \Delta t \rangle = \frac{t_H - t_\mathrm{cut}}{\ln(t_H / t_\mathrm{cut})}
\end{equation}
Thus, changing $t_\mathrm{cut}$ by an order of magnitude from 10 Myr to 100 Myr changes the mean time delay by only 0.17 dex.  As discussed below, the power law index $n$ has a much more significant effect on the mean delay time.  We therefore fix $t_\mathrm{cut}$ to 50 Myr.

In this paper, we consider three possible power-law indices: $n=-1.5$, $n=-1$, and $n=-0.5$.  As noted above, indices with $n \approx -1$ are favored in the literature (c.f.\ \citealt{Hao13}), but we include $n=-1.5$ and $n=-0.5$ to show the insensitivity of our main results on the hostless fraction to the time delay distribution.  The implied cosmic sGRB rates for the different delay-time models are shown in Fig.\ \ref{f:grb_rate}, which are obtained from convolving Eq.\ \ref{e:prob} with the cosmic star formation rate (CSFR).  These rates do not depend on the halo model assumed (only on the cosmic SFR and the time-delay distribution).  However, we note that the overall normalization is uncertain due to the unknown average beam opening angle---and therefore, average detection probability---for sGRBs.   If the delay distributions are truncated at the Hubble time, the average delay times for these distributions are $0.8$, $2.4$, and $4.9$ Gyr, for $n$ of $-1.5$, $-1$, and $-0.5$, respectively; the medians are $0.18$, $0.8$, and $3.9$ Gyr, respectively.

\subsection{Results}

\label{s:halo_results}

We show results for the probability distribution of sGRBs as a function of halo mass and cosmic time, $P(M_h,t)$, in the left-hand panels of Fig.\ \ref{f:grb_halos} for all time-delay distributions considered.  Regardless of the time-delay distribution assumed, the host halos of most sGRBs are in the range $10^{11.5}$ to $10^{12.5}\Msun$.  This is due to the equally narrow range in halo masses where most \textit{star formation} in the universe takes place \citep{BehrooziEvolution}.

The largest visible effect of a change in the time-delay distribution is a change in when the sGRBs occur (see also Fig.\ \ref{f:grb_rate}).  There is a small secondary effect on the median halo mass for sGRBs at late times; because halos grow over time, shorter time-delay distributions will result in lower host halo masses for sGRBs.  This effect is only apparent at $z=0$, where the median sGRB halo mass for the shortest time-delay distribution ($10^{11.7}\Msun$) is about 0.4 dex less than the median sGRB halo mass for the longest time-delay distribution  ($10^{12.1}\Msun$); the ratio of the median stellar masses at $z=0$ for these two delay distributions is very similar (see \S \ref{s:smf_predictions}).

We also show the sGRB rate per unit halo in the right-hand panels of Fig.\ \ref{f:grb_halos}.  Since larger halos host larger galaxies, the sGRB rate almost always increases with halo mass.  Conversely, there is a falloff in the sGRB probability density for halos below about $10^{11.5}\Msun$ in the right-hand panels of Fig.\ \ref{f:grb_halos}.  The main reason for the drop in the probability density for halos above $10^{12.5}\Msun$ (left-hand panels) is that the \textit{number density} of halos falls off rapidly towards larger masses.  Even though the sGRB rate per unit halo is higher, massive halos are too rare to host a significant fraction of sGRBs.

\section{Constraints on the Short GRB Progenitor Velocity Distribution from Observed sGRB--Galaxy Offsets}

\label{s:hostless}

We discuss methods for calculating hostless fractions in \S \ref{s:vel_methods}, predictions for observed hostless fractions for several progenitor velocity distributions in \S \ref{s:hostless_results}, and the robustness of our results to observational and instrumental systematics in \S \ref{s:systematics}.

\subsection{Methods}

\label{s:vel_methods}

\begin{figure}
\plotgrace{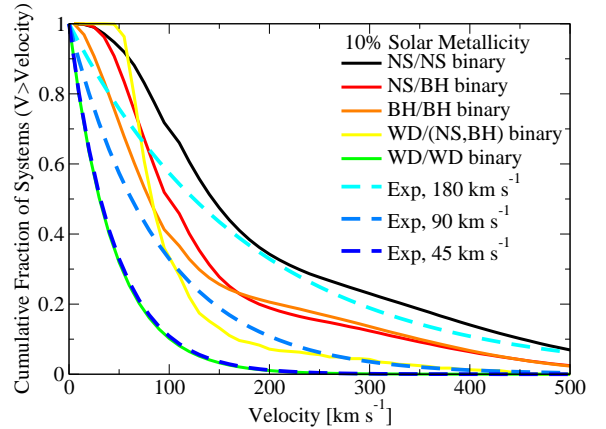}
\caption{Cumulative velocity offset probability distributions for several types of potential sGRB progenitors, generated from population synthesis models.  These include mergers between any combination of neutron stars (\textbf{NS}), black holes (\textbf{BH}), and white dwarfs (\textbf{WD}) for populations at 10\% solar metallicity.  \textit{Dashed lines} show exponential distributions for comparison with $\langle v \rangle$ of 180, 90, and 45 km s$^{-1}$.  These velocities include contributions both from natal kicks and from stellar velocity dispersions; the latter is relevant mainly for binary white dwarf mergers.}
\label{f:vel_dist}
\end{figure}

\subsubsection{Escape of Short GRB Progenitors}

\label{s:escape_fraction}

As most sGRBs occur in halos of mass near $10^{12}\Msun$, there are significant kinetic energy requirements for sGRBs to escape from their host galaxies.  As an example, traveling from the halo center to $0.1\rvir$ in a $10^{12}\Msun$ NFW \citep{NFW97} halo requires an initial velocity above 300 km s$^{-1}$.  The observed sGRB hostless fraction, in combination with priors on the host halo masses of sGRBs from $\S \ref{s:halo_masses}$, therefore constrains the sGRB progenitor velocity distribution.

We have estimated the velocity kick distributions for binary combinations of neutron stars, black holes, and white dwarfs using population synthesis code \citep{Fryer98,Fryer99}.  Based on the pulsar velocity distribution \citep{Arzoumanian02}, we assume a bimodal velocity distribution for kicks imparted onto newly formed neutron stars.  Black holes formed with no accompanying supernova explosion are assumed to receive no kick.  Black holes formed with weak supernova explosions and considerable fallback are assumed to have similar kick momenta as that received by neutron stars.  However, due to their higher masses, the black hole kick velocities are less than those for neutron stars.  For white dwarf/white dwarf binaries, the velocity distribution is set to the local stellar velocity dispersion \citep{Binney08}.  White dwarf/black hole and white dwarf/neutron star systems all have some minimum kick ($\approx$ 60 km s$^{-1}$) in systems that remain bound after the formation of the neutron star or black hole, but the fraction of these systems with kicks above 200 km s$^{-1}$ is small.
White dwarf---neutron star merger timescales are on the order of minutes and so have primarily been considered as long GRB progenitors \citep{Fryer99b}, yet we include these systems as they may be the progenitors of extended emission sGRBs \citep{King07,Lee10,Norris10} or the precursors of accretion-induced collapse in neutron stars \citep{Fryer99c,Giacomazzo12}.  Fig.\ \ref{f:vel_dist} shows the cumulative fraction of each compact binary system as a function of minimum velocity.
 
Since most stars form in $\approx 10^{12}\Msun$ halos, the high-velocity tail ($>200$ km s$^{-1}$) of the velocity distribution has the largest impact on the observed hostless fraction.  Because the exact functional form for the progenitor distribution is unknown, we test three different exponential velocity distributions, $P(v) = \exp(-\lambda v)$, with mean velocities $\langle v \rangle$ of 45, 90, and 180 km s$^{-1}$.  As shown in Fig.\ \ref{f:vel_dist}, these distributions approximately bracket the high-velocity tails of the different progenitor classes.  Binary white dwarf mergers correspond to the 45 km s$^{-1}$ distribution; white dwarf / neutron star or white dwarf / black hole mergers correspond approximately to the 90 km s$^{-1}$ distribution; binary neutron star mergers correspond approximately to the 180 km s$^{-1}$ distribution; and neutron star / black hole mergers fall in between the 90 and 180 km s$^{-1}$ distributions.  We have also considered Maxwell-Boltzmann velocity distributions (Appendix \ref{a:maxwell}) and verified that our main conclusions are unchanged by the alternate functional form.  Additionally, we discuss uncertainties in the neutron star progenitor kick velocities in Appendix \ref{a:ns_uncertainties}.

We then employ a simple, yet conservative model for estimating the upper bound on the observed hostless fraction.  Given a choice of time-delay distribution (\S \ref{s:halo_masses}), a velocity distribution for sGRB progenitors, and the corresponding host halo masses, we can calculate the distribution of radii which the progenitors reach before the sGRB occurs.  We assume that the sGRB progenitors begin at the half-light radii of their galaxies (well-approximated by $0.015 R_\mathrm{200c}$; \citealt{Kravtsov13}),\footnote{$R_\mathrm{200c}$ refers to the radius from the halo center within which the average over-density is 200 times the critical density.} and we assign a purely radial velocity chosen from the assumed velocity distribution.  While progenitor velocities in the real universe will have a tangential component, assigning a purely radial velocity ensures that the model gives an upper bound on the hostless fraction (see also Appendix \ref{s:tests}).

The term ``hostless'' \citep[e.g.,][]{Berger10,Fong13,Fong13b} does not directly specify the distance between the sGRB and the progenitor galaxy.  We consider three separate thresholds $r_\mathrm{thresh}$ for where sGRBs would be considered ``hostless''---i.e., $5 \Reff$, $10 \Reff$, and $15 \Reff$ (where $\Reff$ is the host galaxy half-light radius), to explore the expected radial distribution relative to the true galaxy host.  We determine upper bounds on the hostless population by calculating the fraction of sGRB progenitors which receive velocity kicks large enough to reach these threshold radii, as measured in 3D.  This ignores the fact that the sGRB progenitors will spend time below the turnaround radius; also, it ignores the fact that the projected 2D radius will always be less than the true 3D radius.  Both of these effects will reduce the true hostless fraction below our estimated upper bounds.  For simplicity, we also ignore the (weak) influence of dark energy on the gravitational force within halos, as well as the changing potential due to halo mass accretion.  Appendix \ref{s:tests} tests the appropriateness of ignoring these effects, as well as our other assumptions, with more realistic simulations.  The results of Appendix \ref{s:tests} confirm that the hostless fractions calculated by the method in this section provide a slight overestimate of the true hostless fractions.  Finally, we ignore the effects of galaxy-galaxy mergers, as simulations suggest that $<0.1\%$ of stars can be unbound from the central galaxy \citep{BehrooziUnbound} in these collisions.

We summarize our escape model as follows:
\begin{enumerate}
\item For a given threshold radius $r_\mathrm{thresh}$, we calculate the minimum radial velocity ($v_r$) necessary for a particle to travel from $0.015 R_\mathrm{200c}$ to $r_\mathrm{thresh}$ in a NFW halo potential for a wide range of halo masses ($10^{7}\Msun < M_h < 10^{16}\Msun$) and redshifts ($0 < z < 8$).
\item For a given initial velocity distribution for sGRB progenitors, we calculate the fraction of sGRBs which have speeds $|v|$ greater than this minimum radial velocity, also as a function of halo mass and redshift.
\item We weight these fractions (i.e., $P(|v| > v_r | M_h, z)$) by the probability distribution for halo masses and redshifts of sGRBs, $P(M_h, z)$, for a given time-delay distribution (\S \ref{s:halo_masses}), to calculate an upper bound for the total hostless fraction.
\end{enumerate}

\subsubsection{Instrumental, Geometrical, and Observational Biases}

\label{s:biases}

In \S \ref{s:halo_masses}, we calculated the volume density of sGRBs as a function of host halo mass and redshift.  However, the observed redshift distribution for sGRBs will be biased due to geometrical and instrumental effects:
\begin{equation}
\label{e:sgrbr_obs}
sGRBr_\mathrm{obs} = \int_0^{\infty} sGRBr_\mathrm{intrinsic}(z) \frac{dV(z)}{dz} \frac{1}{1+z}  \epsilon_\mathrm{detector}(z) dz,
\end{equation}
where $sGRBr_\mathrm{obs}$ is the observed sGRB rate (per unit time), $sGRBr_\mathrm{intrinsic}(z)$ is the intrinsic sGRB rate per unit time per unit volume (calculated in \S \ref{s:halo_masses}), $V(z)$ is the comoving volume out to redshift $z$, $(1+z)$ is the time dilation factor, and $\epsilon_\mathrm{detector}(z)$ is the mean detection probability for sGRBs at redshift $z$.

The redshift sensitivity, $\epsilon_\mathrm{detector}(z)$, is constrained both by the apparent luminosity function of sGRBs (e.g., from BATSE or Fermi; \citealt{Goldstein12,Goldstein13}) and by the observed redshift distribution (e.g., \citealt{Fong13}).  Since the deconvolution process is nontrivial \citep{Guetta06}, we present our analysis in Appendix \ref{a:sensitivity}; for the \textit{Swift} BAT, we find (Eq.\ \ref{e:detector_fit_final}):
\begin{equation}
\label{e:detector}
\epsilon_\mathrm{detector}(z) \propto \exp(-4.3z),
\end{equation}
with an unknown overall constant of proportionality dependent on the unknown beaming angles of sGRBs.

\begin{figure}
\plotgrace{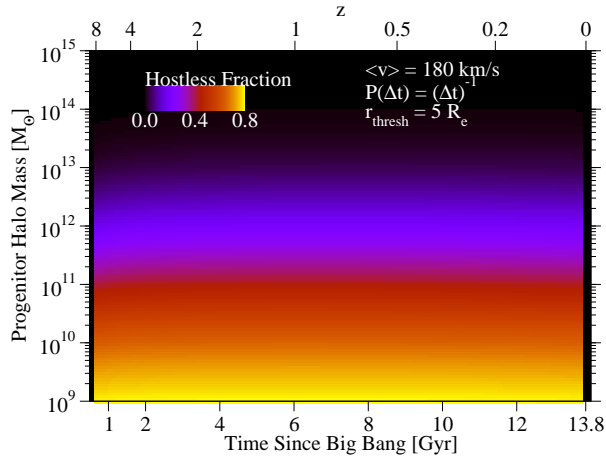}
\caption{The fraction of hostless sGRBs per galaxy as a function of the galaxy's host halo mass and redshift (see \S \ref{s:escape_fraction}), for an example model.  This model assumes a mean progenitor kick velocity of $\langle v\rangle=180$ km s$^{-1}$ (with an exponential velocity distribution; $P(v) = \exp[-v/\langle v \rangle]$), a power-law delay time distribution with $P(\Delta t) \propto (\Delta t)^{-1}$, and the threshold for being considered ``hostless'' placed at 5 times the host galaxy's effective radius.  Very little redshift evolution is seen in this or other models; for a fixed velocity distribution, the host halo mass is the primary determinant of escape.}
\label{f:grb_ef}
\end{figure}

\begin{figure}
\plotgrace{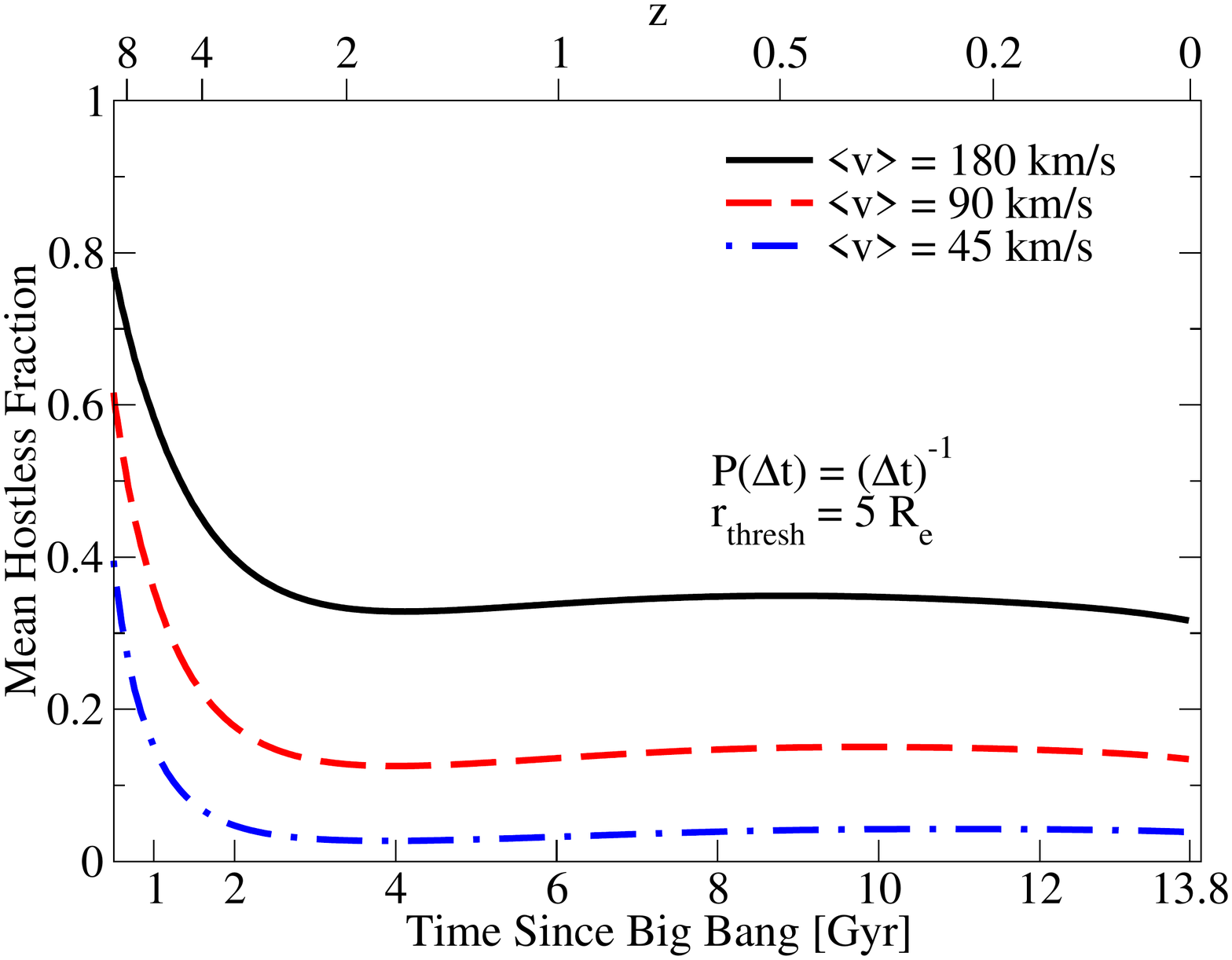}\\[-4ex]
\plotgrace{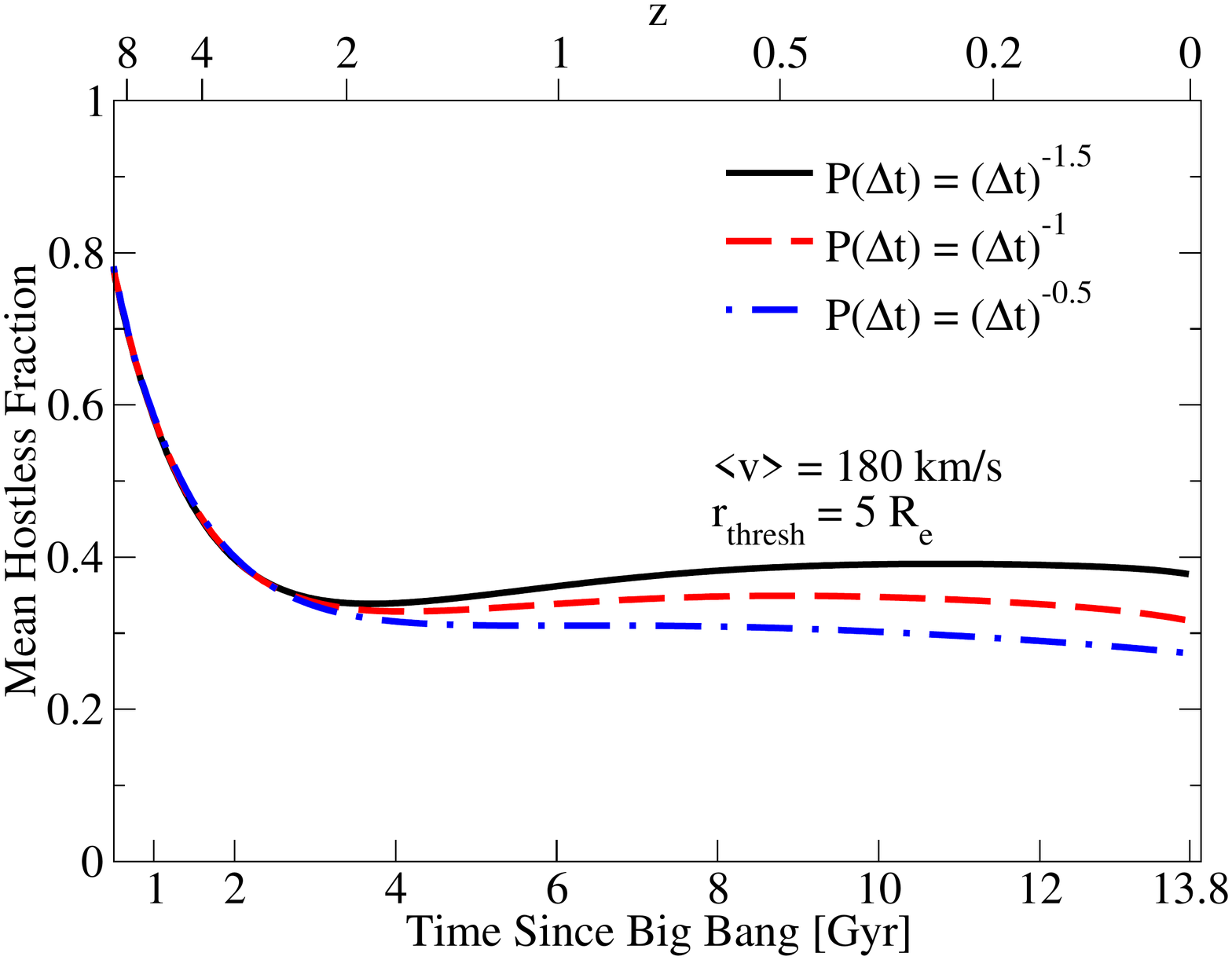}\\[-4ex]
\plotgrace{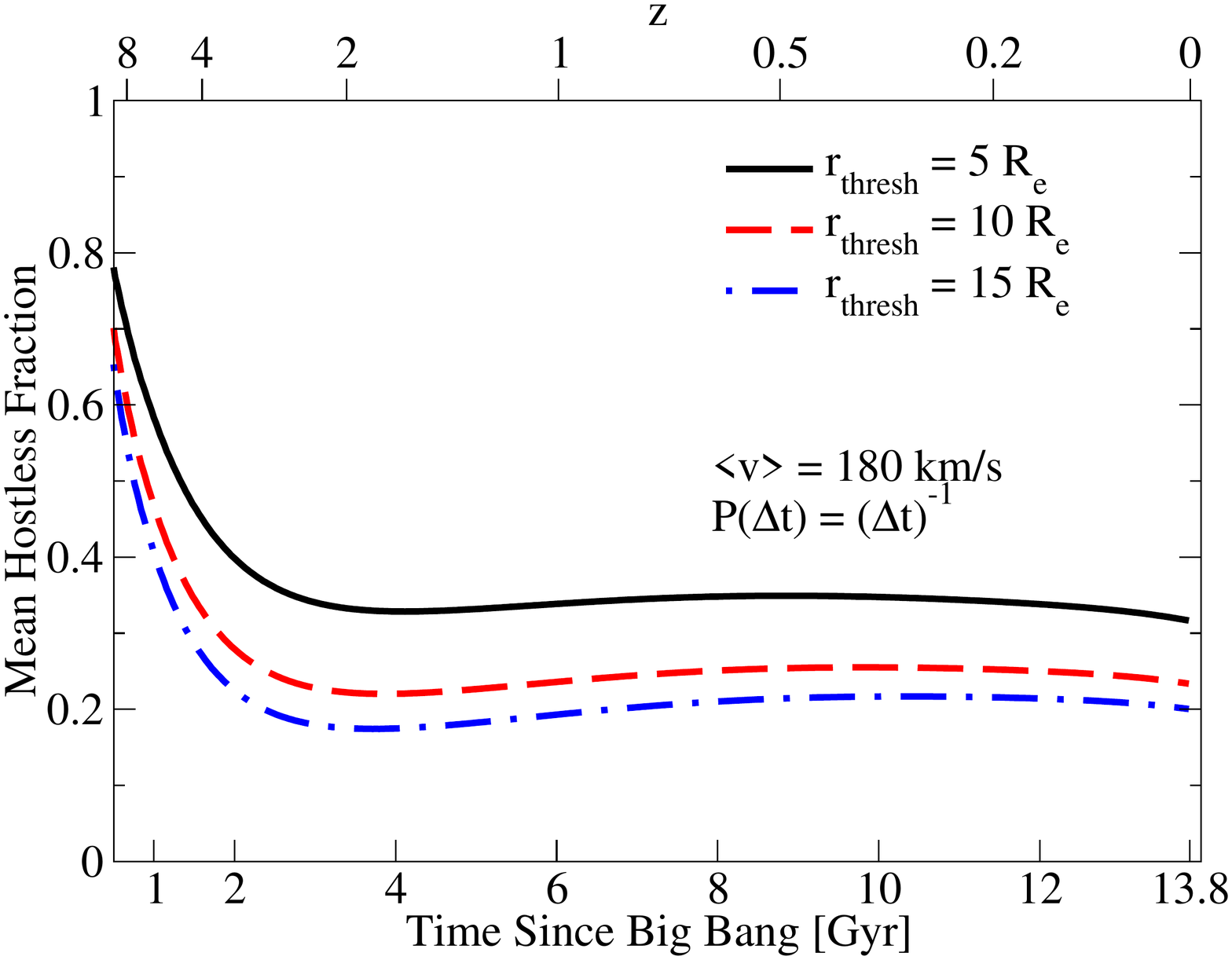}\\[-4ex]
\caption{\textbf{Top Panel}: the fraction of hostless sGRBs as a function of redshift and velocity distribution, holding fixed the power-law delay time distribution ($P(\Delta t) \propto (\Delta t)^{-1}$) and the threshold for being considered ``hostless'' (5 times the host galaxy's effective radius). \textbf{Middle Panel}: same, but instead varying the time delay distribution while holding the velocity distribution and hostless threshold fixed ($\langle{}v\rangle$ = 180 km s$^{-1}$ and $r_\mathrm{thresh} = 5\Reff$, respectively).   \textbf{Bottom Panel}: same, but varying the hostless threshold while holding the velocity distribution and time delay distribution ($\langle v \rangle$ = 180 km s$^{-1}$ and $P(\Delta t) \propto (\Delta t)^{-1}$, respectively).
Because the halo mass where most stars form is also constant over much of cosmic history \citep{BehrooziEvolution}, there is very little evolution in the mean hostless fraction of sGRBs from $z=2$ to $z=0$.  At $z>2$, the evolution of the halo mass function results in fewer high-mass halos being available to host galaxies, increasing average escape fractions.  However, most star formation occurred at $z<2$, meaning that reweighting the contribution from sGRBs at different redshifts does not change the observed hostless fraction.  This limits the impact of many systematic uncertainties, including the \textit{Swift} redshift sensitivity.}
\label{f:grb_ef_models}
\end{figure}

The fraction of observed sGRBs which are truly hostless is remarkably insensitive to the redshift sensitivity in Eq.\ \ref{e:detector} (see also \S \ref{s:systematics}).  The median halo mass where stars form has remained almost exactly the same since $z=4$ \citep{BehrooziEvolution}, as has the median halo mass where sGRBs occur (Fig.\ \ref{f:grb_halos}, left panels).  The energy required to escape from a given halo mass also does not evolve over this time (Fig.\ \ref{f:grb_ef}).  Therefore, for a given progenitor velocity distribution, the escape fraction does not change significantly as a function of the sGRB's redshift (Fig.\ \ref{f:grb_ef_models}).

That said, $\epsilon_\mathrm{detector}(z)$ can affect the fraction of sGRBs with unobservable hosts.  Short GRBs at higher redshifts are more likely to have faint galaxy hosts, which would artificially increase the observed hostless fraction if they were missed in the followup observations.  For this reason, we also examine the effects of significantly extending \textit{Swift}'s redshift sensitivity in \S \ref{s:systematics}.  On the other hand, faint galaxies are more likely to reside in small halos with shallow potential wells, meaning that many sGRBs with faint progenitor galaxies are more likely to be hostless to begin with (see, e.g., Fig.\ \ref{f:grb_ef}).

To estimate the net change in the hostless fraction due to limited-depth followup observations, it is necessary to calculate host galaxy luminosities as a function of host halo mass and redshift.  We do so by convolving galaxy star formation histories as a function of halo mass and redshift \citep{BWC13} with the FSPS stellar population synthesis model \citep{Conroy09,Conroy10} to obtain galaxy luminosities in the Hubble WFC3 F160W band.  In this calculation, we have assumed the metallicity of stars formed at a given redshift tracks the gas-phase metallicity in \cite{Maiolino08}, and have adopted a dust optical depth of $\tau=0.1$ (calibrated to match CANDELS observations of the stellar mass -- F160W relation from $z=0.6$ to $z\approx6$, for $M_\ast < 10^{10}\Msun$; \citealt{Chang13,Salmon13}).  Given host galaxy luminosities estimated in this way, it is straightforward to calculate how many observed sGRBs would not have detectable hosts in limited-depth followup observations. For this purpose, we adopt the same host galaxy detection threshold of F160W$<$26 mag (AB) as in \cite{Fong13b}.

Finally, we note that the reported fraction of sGRBs at large radial offsets \citep{Fong13b} is almost certainly a lower limit for the true distribution.  Obtaining a precise position with \textit{Swift} requires an afterglow; however, merging binaries are expected to show an environmental dependence in their afterglow signatures \citep{Panaitescu01,Lee05}.  At $>5\Reff$, the external shock would take place within a very tenuous medium, implying a bias against detecting afterglows for highly-offset sGRBs \citep{Salvaterra10}.  The true fraction of sGRBs taking place at this distance could therefore be as large as 50\%, if this effect were responsible for all sGRBs with XRT followup that lack sub-arcsec positions \citep{Fong13}.

\subsection{Results}
\label{s:hostless_results}
\begin{figure}
\plotgrace{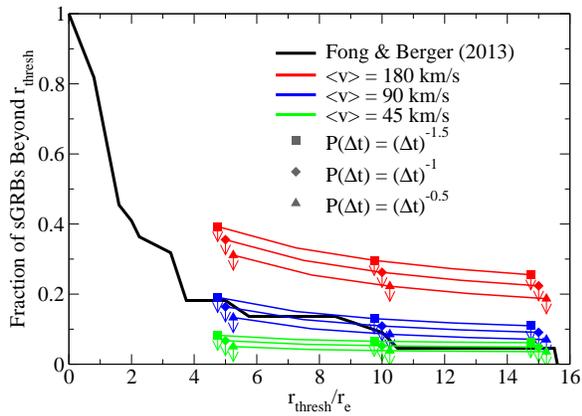}
\caption{A comparison between the observed locations of sGRBs from \cite{Fong13b} and the predicted locations of sGRBs from Table \ref{t:hostless_fracs} for several assumptions about the sGRB progenitor kick velocities and sGRB merger time delays.  The influence of the time delay assumption is minimal; the different models have therefore been slightly offset for clarity.  The influence of the velocity kick is more significant; however, due to the assumptions in our model (\S \ref{s:escape_fraction}), we only place upper limits on the fraction of sGRBs which can escape to a given distance.}
\label{f:fong_comp}
\end{figure}

\begin{figure*}
\plotgrace{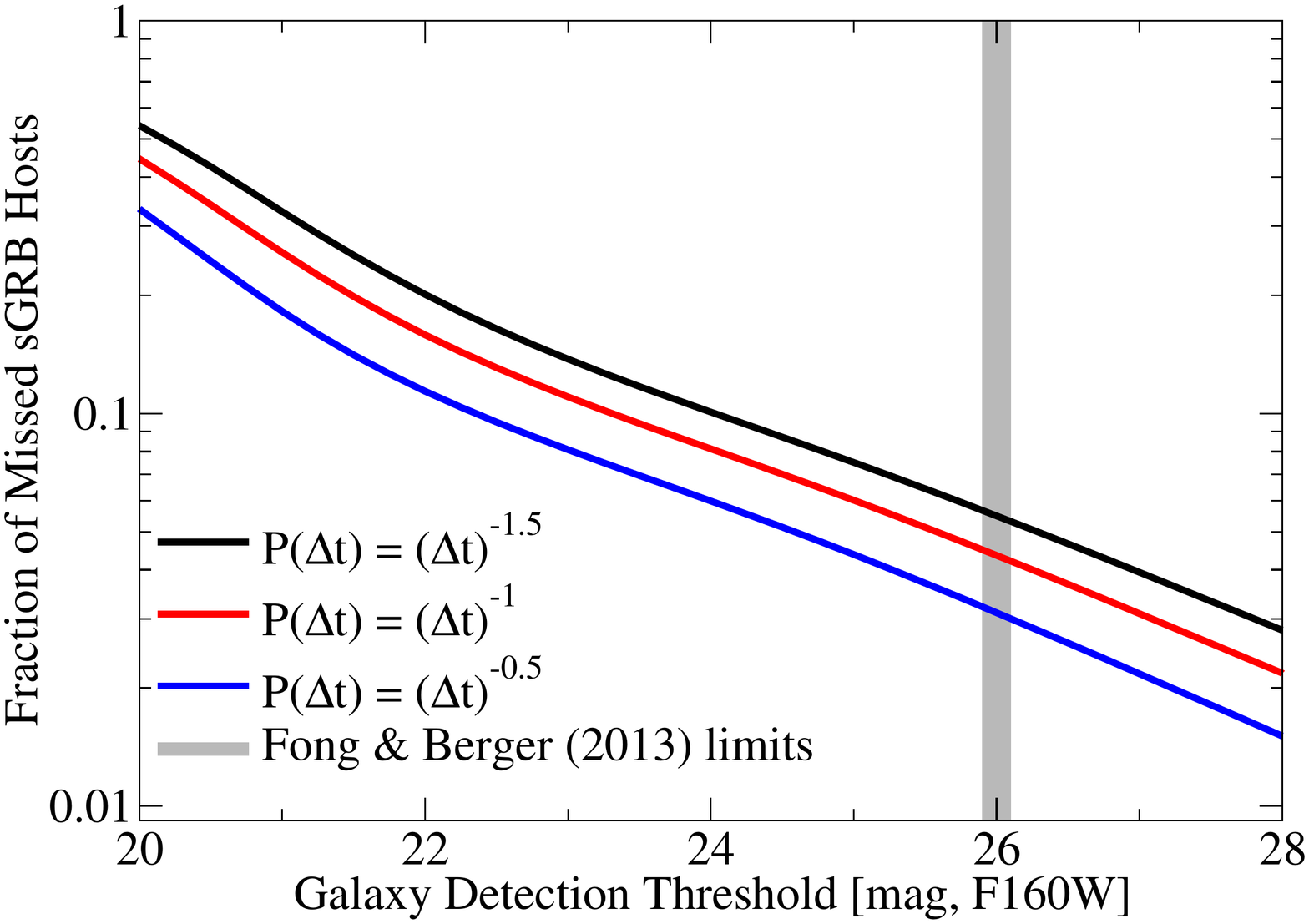}\plotgrace{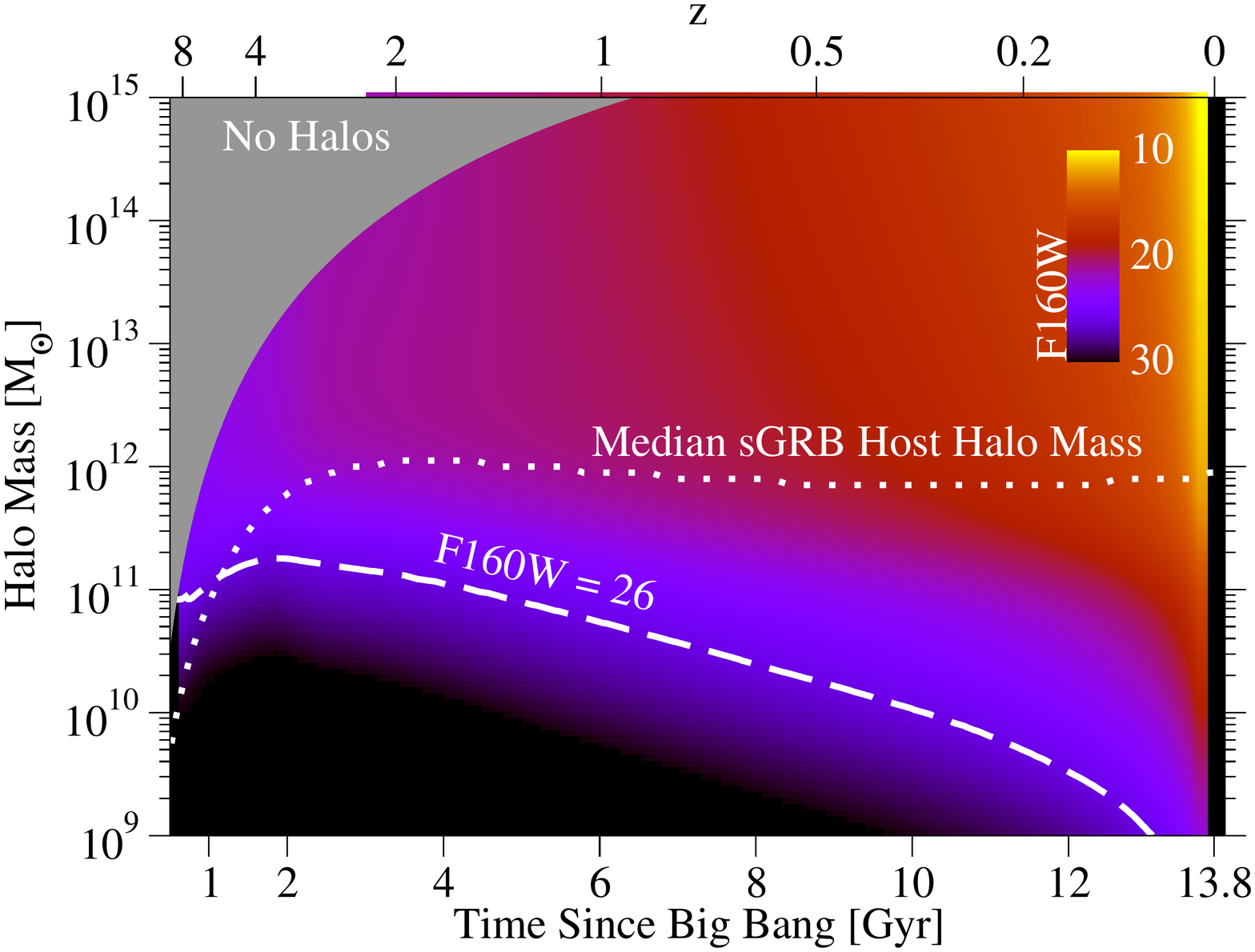}
\caption{\textbf{Left panel}: expected fraction of missed \textit{Swift}-detected sGRB hosts, as a function of the depth of follow-up observations in the F160W band (AB).  The shaded region shows typical limits from \citep{Fong13}, which are F160W $<$ 26 mag.  Only a fraction of sGRB hosts are missed; these correspond mainly to faint galaxies in shallow potential wells---i.e., where the sGRB progenitors are likely to have escaped their host galaxy anyway.  \textbf{Right panel}: Average galaxy luminosity in the F160W band as a function of galaxy host halo mass and redshift.  The \textit{dotted line} shows the median halo mass in which sGRBs appear as a function of redshift, for a $P(\Delta t) \propto (\Delta t)^{-1}$ time-delay distribution (from Fig.\ \ref{f:grb_halos}).  The \textit{dashed line} shows the host halo mass probed by observations which reach F160W $<$ 26 mag (AB).  Such observations are able to detect most sGRB progenitor host galaxies out to $z\approx 4$.}
\label{f:grb_missed_hosts}
\end{figure*}

\begin{table}
\begin{center}
\caption{sGRB Hostless Fractions}
\label{t:hostless_fracs}
\begin{tabular}{ccccc}
\hline
\hline
$\langle v \rangle$ (km s$^{-1}$) & n & $f(r>5\Reff)$ & $f(r>10\Reff)$ & $f(r>15\Reff)$ \\
\hline
45 & -0.5 & <0.050 & <0.039 & <0.036\\
45 & -1.0 & <0.067 & <0.053 & <0.049\\
45 & -1.5 & <0.082 & <0.065 & <0.061\\
\hline
\textbf{90} & \textbf{-0.5} & <\textbf{0.133} & <\textbf{0.085} & <\textbf{0.070}\\
\textbf{90} & \textbf{-1.0} & <\textbf{0.164} & <\textbf{0.109} & <\textbf{0.091}\\
\textbf{90} & \textbf{-1.5} & <\textbf{0.191} & <\textbf{0.130} & <\textbf{0.110}\\
\hline
\textbf{180} & \textbf{-0.5} & <\textbf{0.311} & <\textbf{0.223} & <\textbf{0.187}\\
\textbf{180} & \textbf{-1.0} & <\textbf{0.355} & <\textbf{0.262} & <\textbf{0.224}\\
\textbf{180} & \textbf{-1.5} & <\textbf{0.393} & <\textbf{0.296} & <\textbf{0.255}\\
\hline
\end{tabular}
\end{center}
\tablecomments{Upper limits on the observed fraction of sGRBs which have escaped their host galaxies, for a range of different assumptions for the mean progenitor kick velocity ($\langle v\rangle$; \S \ref{s:vel_methods}) and the time-delay power law index ($n$; \S \ref{s:halo_masses}).  $f(r>5\Reff)$ refers to the fraction of sGRBs which occur at distances greater than 5 times the effective radius of their host galaxies; $f(r>10\Reff)$ and $f(r>15\Reff)$ denote the fractions which occur beyond 10 and 15 $\Reff$, respectively.  Several systematics, including \textit{Swift} sGRB detectability and limited-depth optical/infrared followup are included, although these do not materially affect the results (see \S \ref{s:systematics}).  \textbf{Bold} letters indicate combinations which result in hostless fractions consistent with \cite{Fong13} and \cite{Fong13b}.  These results strongly suggest that the mean progenitor kick velocity is greater than 45 km s$^{-1}$.}
\end{table}

We show results for observed hostless fraction upper bounds in Table \ref{t:hostless_fracs}, and show the comparison to \cite{Fong13b} in Fig.\ \ref{f:fong_comp}.  Regardless of the initial assumptions for the time-delay model or hostless threshold radius, hostless fractions for an exponential velocity distribution with $\langle v \rangle = 45$ km s$^{-1}$ are all $<9\%$, well below the hostless fraction of 18\% for sGRBs occuring beyond 5$\Reff$ suggested by \cite{Fong13b}.  With the small sample size in \cite{Fong13b} (22 sGRBs), the 45 km s$^{-1}$ models are excluded with confidence levels of 90 -- 98\%.  \cite{Fong13} report a hostless fraction of 17\% for 36 sGRBs (which overlap with the sample in \citealt{Fong13b}), although not all of these have sub-arcsecond locations; for this sample, the 45 km s$^{-1}$ models would be excluded with confidence levels between 93 -- 99.2 \%.

The results in Table \ref{t:hostless_fracs} show the largest variation with the choice of velocity distribution, significantly smaller variation with the threshold radius, and finally, least variation with the delay-time model (see also Figs.\ \ref{f:grb_ef_models} and \ref{f:fong_comp}).  As shown in \S \ref{s:halo_masses}, changing the delay-time model has only a small effect on the host halo masses.  Longer delay times result in slightly larger halo masses at late times, which makes it modestly harder for sGRB progenitors to escape from their host galaxies.  Changing the threshold radius has a larger impact; it takes approximately 1.8 times as much energy to travel to $10 \Reff$ as it does to reach $5 \Reff$, for example.  However, doubling the mean velocity from 90 km s$^{-1}$ to 180 km s$^{-1}$ quadruples the amount of kinetic energy available, which has the largest effect in Table \ref{t:hostless_fracs} and in Fig.\ \ref{f:fong_comp}.

From the exponential distribution with limits closest to the \cite{Fong13b} results ($\langle v \rangle = 90$ km s$^{-1}$), we expect that at least one in five sGRB progenitors received a natal kick larger than $150$ km s$^{-1}$ (Fig.\ \ref{f:vel_dist}).  We discuss the impact of this result on sGRB progenitor models, compare with previous velocity kick estimates \citep{Fong13b}, and provide constraints on arbitrary velocity distributions in \S \ref{s:progenitors}.

\subsection{Robustness to Observational and Instrumental Systematics}

\label{s:systematics}

The vast majority of the systematic effects discussed in \S \ref{s:biases} can be considered as reweighting the contributions from sGRBs at different redshifts; these include the change in observable comoving volume as a function of redshift, time dilation as a function of redshift, and the \textit{Swift} detector sensitivity.  As shown in Table \ref{t:hostless_fracs_syst} (Appendix \ref{s:quant}), reweighting the redshift distribution of sGRBs affects the observed hostless fractions at the 1-2\% level.  As noted in \S \ref{s:biases}, this is primarily because most stars are created at $z\approx2$ and are formed in a narrow range of host halo masses near $10^{12}\Msun$ \citep{BehrooziEvolution}.  As this mass scale determines the energy required to escape independent of redshift, the hostless fraction of sGRBs is also a very weak function of redshift (see, e.g., Figs.\ \ref{f:grb_ef} and \ref{f:grb_ef_models}).

The effects of limited-depth follow-up observations are important only for the 45 km s$^{-1}$ velocity distribution, as shown in the ``Actual Obs.'' column of Table \ref{t:hostless_fracs_syst}.  This column adds the effects of follow-up observations to F160W $<$ 26 mag (AB) in the host galaxy luminosity, matching the limit in \cite{Fong13b}.  Fig.\ \ref{f:grb_missed_hosts} shows the fraction of missed galaxy hosts for \textit{Swift}-detected sGRBs as a function of the limiting luminosity and the time-delay distribution; it also shows the average F160W luminosities of sGRB host galaxies as a function of halo mass and cosmic time.  The limit of F160W $<$ 26 mag is deep enough that the galaxies which contribute most of the sGRB population in the universe are captured out to at least $z\approx 4$ (Fig.\ \ref{f:grb_missed_hosts}, right panel).  Since F160W luminosity correlates strongly with halo mass, galaxies living in shallow halo potential wells are more likely to be missed; these are also the galaxies from which the sGRB progenitors are most likely to escape.  This explains why the fraction of missed hosts in Fig.\ \ref{f:grb_missed_hosts} (left panel) is not additive with the fraction of hostless sGRBs in the ``True for \textit{Swift} Obs.'' column in Table \ref{t:hostless_fracs_syst} (i.e., the hostless fraction for infinitely deep follow-up observations).  For example, the observed hostless fractions for the 180 km s$^{-1}$ models are affected only at the 0--2\% level.    However, the fraction of missed hosts in Fig.\ \ref{f:grb_missed_hosts} does set a lower floor of 3-6\% for the observed ``hostless'' fraction; this is evident in the boosted hostless fraction for all the $\langle v \rangle = 45$ km s$^{-1}$ models in Table \ref{t:hostless_fracs_syst} with finite-depth follow-up.

Finally, the fact that most sGRB hosts would be captured out to $z\approx 4$ with the current depth of follow-up observations means that these results are very insensitive to the actual \textit{Swift} redshift-dependent detection probability.  The ``Upgraded \textit{Swift}'' column in Table \ref{t:hostless_fracs_syst} gives the expected hostless fraction if \textit{Swift}'s redshift-dependent sensitivity were upgraded to $\epsilon_\mathrm{detector} = \exp(-z)$.  This sensitivity is far beyond any expectation of its current abilities \citep{Guetta06,Coward12,Kelley13}.  Even with the same limited-depth follow-up observations (F160W $<$ 26 mag), the expected observed hostless fractions change only at the 1-2\% level for ``Upgraded \textit{Swift}'' (Table \ref{t:hostless_fracs_syst}).

\begin{figure}
\plotgrace{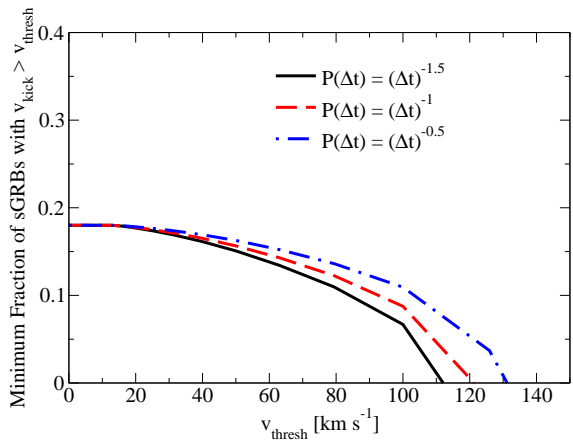}
\caption{Minimum fraction of sGRB natal kicks required to be above a given velocity threshold, $v_\mathrm{thresh}$, in order to match sGRB---galaxy offsets in \cite{Fong13b}.  These minima are valid for \textit{arbitrary} kick velocity distributions; see \S \ref{s:progenitors} and Eq.\ \ref{e:min_vel}.  These limits are a weak function of the assumed time delay model; longer time delays imply slightly higher velocity minima.}
\label{f:grb_escape_energy}
\end{figure}

\section{Discussion}
\label{s:discussion}

We discuss constraints on sGRB progenitor models and natal kick distributions in \S \ref{s:progenitors}, sGRB host demographics and the observable effects of different time-delay models in \S \ref{s:time_delay}, local sGRB host demographics and gravitational wave followup in \S \ref{s:gravitational_waves}, and possible links between sGRBs and globular clusters in \S \ref{s:globular}.

\subsection{Impact of sGRB---Galaxy Offsets on Kick Velocity Distributions and Allowable sGRB Progenitors}

\label{s:progenitors}

\cite{Fong13} and \cite{Fong13b} have found that at least 17--18\% of sGRBs occur beyond five effective radii ($\Reff$) of the nearest likely host galaxy (see discussion at the end of \S \ref{s:biases}).  As noted in \S \ref{s:hostless_results}, it is very difficult for kick velocity distributions with $\langle v \rangle = 45$ km s$^{-1}$ and below to reproduce these sGRB offsets.  White dwarf--white dwarf mergers are therefore strongly disfavored as progenitor candidates.  However, other types of mergers, including white dwarf--neutron star, binary neutron star, and neutron star--black hole mergers would all be allowed by the present limits.

The model estimates in this paper were chosen to be simple and conservative; i.e., to provide stringent upper bounds for the hostless fraction of sGRBs.  As shown in Fig.\ \ref{f:vel_dist}, current models predict a significant energy difference between white dwarf--neutron star velocity kicks and binary neutron star velocity kicks.  So, it may be possible to exclude one of these progenitor classes with more advanced modeling, such as direct injection of tracer particles into halos \citep[e.g.,][]{Zemp07,Kelley10}.  This kind of modeling would also allow recovery of the full \textit{shape} of the progenitor velocity kick distribution from the radial distribution of sGRBs (see also Appendix \ref{a:maxwell}).  We note, however, that the best results for this modeling would also require more precise positions for sGRBs without afterglows.

Based on the match of the 90 km s$^{-1}$ exponential velocity distribution to the offsets reported in \cite{Fong13b}, we estimate that at least 19\% of sGRB progenitors received natal kicks of more than 150 km s$^{-1}$ (Figs.\ \ref{f:vel_dist} and \ref{f:fong_comp}).  These velocities are higher than those reported in \cite{Fong13b}, which were calculated by dividing projected sGRB---host galaxy offsets by the typical ages of the host galaxy stellar populations.  This is to be expected, since the latter method ignores the energy required to leave the host galaxy, as well as the chance for multiple orbits of the progenitors.  We note as well that stellar ages calculated from fitting galaxy luminosities are also extremely uncertain \citep{Pforr12}; this is especially so for those in \cite{Fong13b}, which were derived assuming a single stellar population \citep{Leibler10}.

Our analysis can also place limits on arbitrary velocity distributions.  For any given kick velocity, the method in \S \ref{s:escape_fraction} allows us to calculate the fraction of sGRBs,  $f_5(v)$, that the kick would eject beyond $5\Reff$.  Then, for an arbitrary velocity distribution, $P(v)$, the total ejected fraction, $f$, is given by
\begin{equation}
\label{e:arb_vd}
f = \int_0^\infty P(v)f_5(v)dv.
\end{equation}
Since $f_5(v)$ is monotonically increasing, sGRBs receiving kicks less than a given threshold velocity, $v_\mathrm{thresh}$, can be ejected at most $f_5(v_\mathrm{thresh})$ of the time.  We can then rewrite Eq.\ \ref{e:arb_vd} as a limit on the maximum total fraction of ejected sGRBs:
\begin{equation}
f < P(v<v_\mathrm{thresh}) f_5(v_\mathrm{thresh}) + P(v>v_\mathrm{thresh})
\end{equation}
since kicks above $v_\mathrm{thresh}$ can eject up to 100\% of the binaries they affect.  Alternately, if we know the actual ejected fraction, $f$, we can rewrite the inequality using $P(v<v_\mathrm{thresh}) = 1 - P(v>v_\mathrm{thresh})$ to constrain the velocity distribution:
\begin{equation}
\label{e:min_vel}
P(v>v_\mathrm{thresh}) > \frac{f-f_5(v_\mathrm{thresh})}{1 - f_5(v_\mathrm{thresh})},
\end{equation}
which is valid regardless of the functional form of $P(v)$.  These limits are shown in Fig.\ \ref{f:grb_escape_energy} for $f=0.18$ \citep{Fong13b}; they are a weak function of the time delay model for the same reason discussed in \S \ref{s:biases}.  The velocity distributions which satisfy these limits are far from minimizing the average kick velocity.  E.g., a velocity distribution with 90\% of kicks less than 80 km s$^{-1}$ would have to have the remaining 10\% of kicks at velocities larger than 1000 km s$^{-1}$, resulting in an average kick velocity of 172 km s$^{-1}$.  We find that the velocity distributions which minimize the average kick velocities would have most kicks at 0 km s$^{-1}$ and then 24, 28, or 34\% of kicks at 200 km s$^{-1}$ for time delay models with $P(\Delta t)\propto (\Delta t)^{-1.5}$, $(\Delta t)^{-1}$, and $(\Delta t)^{-0.5}$, respectively.

We note that sGRB position offsets best constrain the high-velocity tails of the kick velocity distribution.  At low kick velocities, the host galaxy's stellar velocity dispersion and radial distribution of star formation become much more important for determining sGRB locations.  As noted above, velocity kick distributions with the same hostless fraction can have significantly different average velocities (see also Appendix \ref{a:maxwell}).  Yet, it is clear that sGRB kick velocities need to extend well beyond the typical stellar velocity dispersions of their host galaxies.  

\begin{figure}
\vspace{-4ex}
\plotgrace{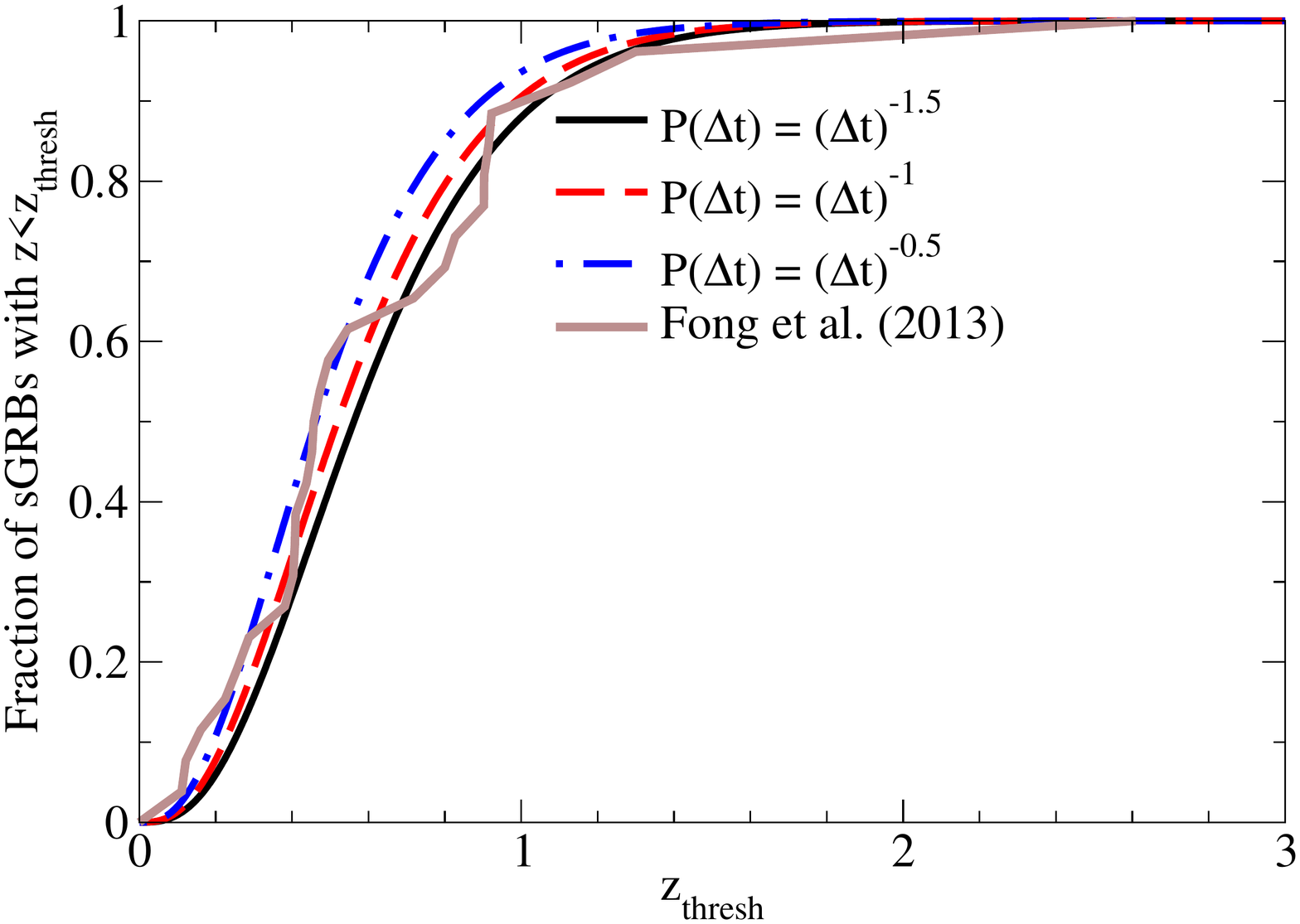}\\[-4ex]
\plotgrace{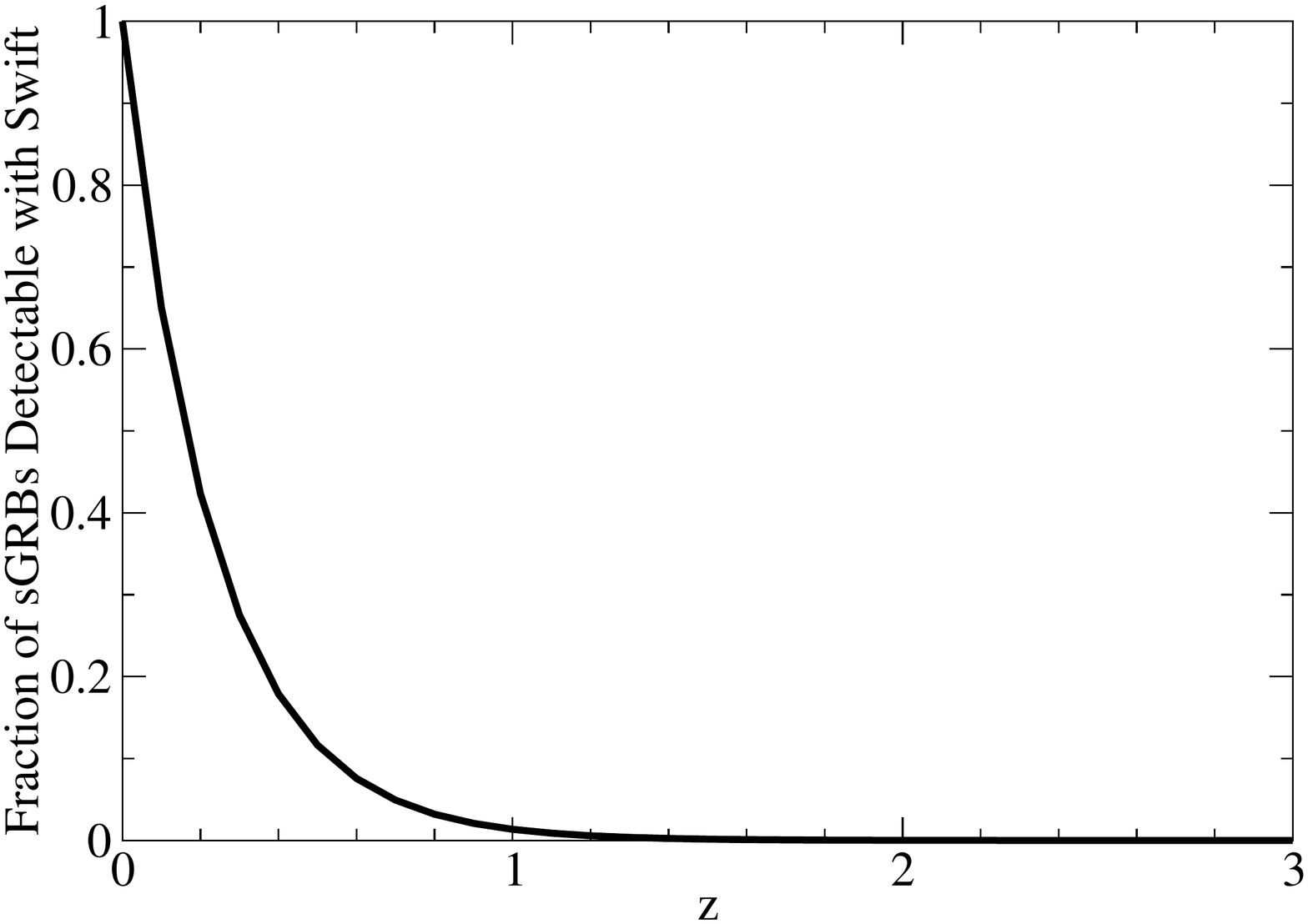}
\caption{\textbf{Top} panel: cumulative observed sGRB redshift distributions for three different time delay models (\S \ref{s:halo_methods}), compared to the observed distribution in \cite{Fong13}.  The differences between time-delay models are significantly reduced compared to Fig.\ \ref{f:grb_rate}, partly because the \textit{Swift} redshift sensitivity is limited for $z > 0.5$ (\textbf{bottom} panel, from Eq.\ \ref{e:detector}), and partly because the observable volume is limited for $z<0.25$.  Hence, the observed redshift distribution from \textit{Swift} does not constrain the time-delay model; Kolmogorov-Smirnov tests show no significant discrepancies ($p>0.48$) between any of the three models and the redshift distribution in \cite{Fong13}.  See Appendix \ref{a:redshift_incomp} for a discussion of how the $\sim$13\% redshift incompleteness for sGRBs with accurate positions may affect these plots.}
\label{f:grb_weighted}
\end{figure}

\begin{figure}
\vspace{-4ex}
\plotgrace{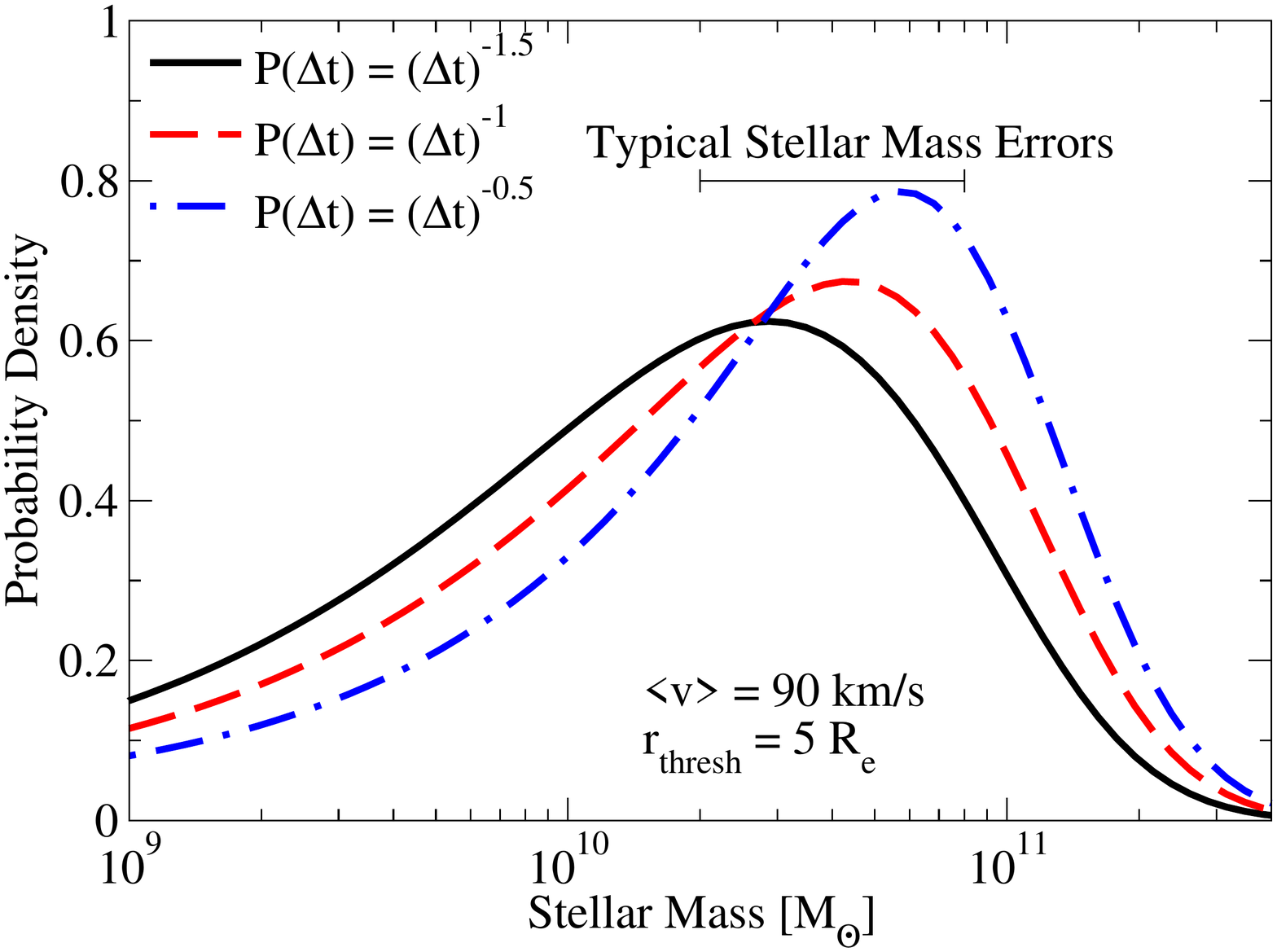}\\[-4ex]
\plotgrace{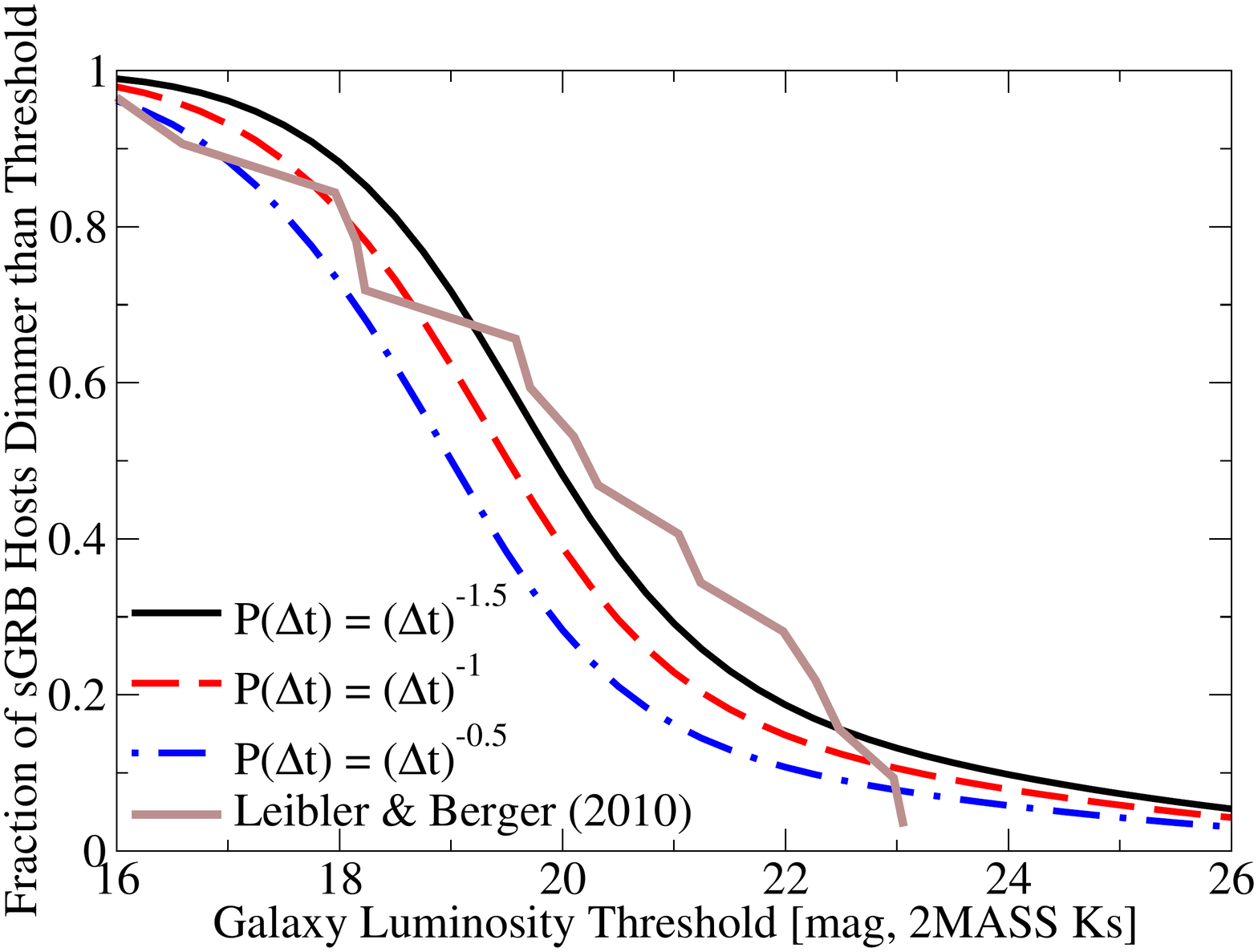}
\caption{\textbf{Top} panel: expected distribution of sGRB host galaxy stellar masses for three different time delay models.  Typical $1\sigma$ systematic biases in stellar masses ($\pm 0.3$ dex) are shown by the horizontal error bar; these exceed the differences present across the three time delay models.  \textbf{Bottom} panel: expected distribution of sGRB host galaxy apparent luminosities (2MASS Ks band, AB magnitudes) for three different time delay models.  Data from \cite{Leibler10} is also shown; Kolmogorov-Smirnov tests do not show inconsistencies with any of the three time-delay models ($p>0.46$ in all cases).}
\label{f:grb_smfs}
\end{figure}

\begin{figure}
\vspace{-4ex}
\plotgrace{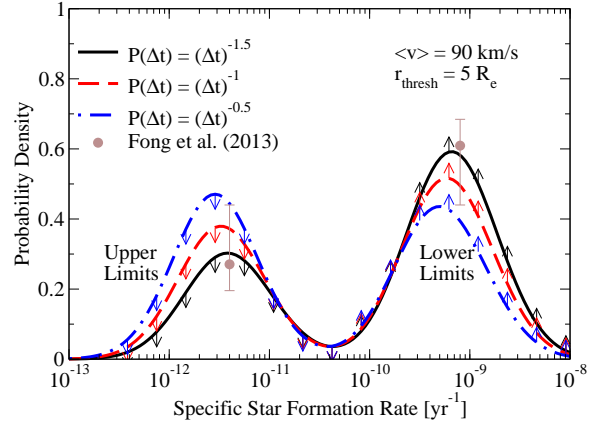}
\caption{Predicted limits for sGRB host galaxy specific star formation rates (i.e., the ratios of star formation rate to stellar mass), for three time-delay models.  The limits for early-type (passive) galaxies are upper limits, since older star formation histories for early-type galaxies will result in them hosting fewer sGRBs (see \S \ref{s:time_delay}).  Similarly, the limits for late-type (active) galaxies are lower limits.  For comparison, data points for the early / late sGRB host fraction are shown from \cite{Fong13}.  The error bars encompass the maximum variation allowed by assigning unclassified sGRB hosts to either all early-type or all late-type galaxies.}
\label{f:grb_ssfrfs}
\end{figure}

\begin{figure}
\vspace{-4ex}
\plotgrace{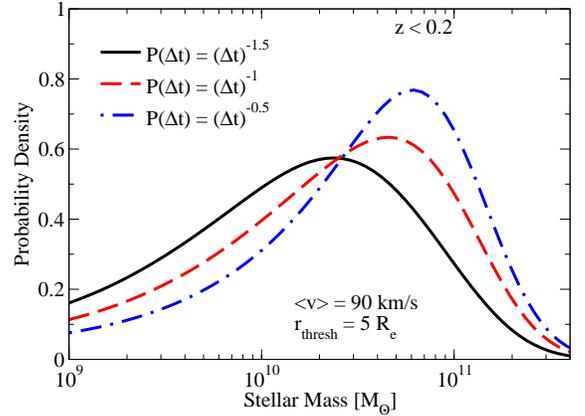}\\[-4ex]
\plotgrace{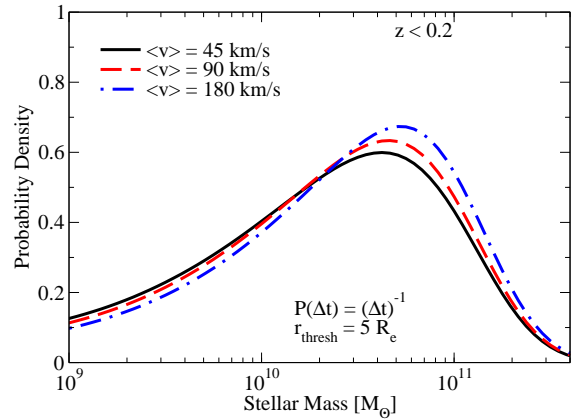}
\caption{Predicted sGRB host galaxy stellar mass probability distribution functions, covering $0<z<0.2$.  \textbf{Top} panel: models where the time-delay distribution is varied and the progenitor kick distribution is held fixed at 90 km s$^{-1}$.  \textbf{Bottom} panel: models where the time-delay distribution is held fixed to $P(\Delta t) \propto (\Delta t)^{-1}$ and the progenitor kick velocities are varied.  In both panels, the threshold radius for being considered ``hostless'' is fixed at 5$\Reff$.}
\label{f:grb_smf}
\end{figure}

\subsection{Determining the Time-Delay Distribution}
\label{s:time_delay}

Constraints on the time-delay distribution of sGRBs have come from the observed redshift distribution of sGRBs \citep{Guetta06,Nakar06}, the host stellar masses \citep{Zheng07GRB,Leibler10}, the fraction of early vs.\ late hosts \citep{Zheng07GRB,Shaughnessy08}, and observational constraints on binary pulsar merger timescales \citep{Kalogera01,Oslowski11}.  In this section, we examine the relative effectiveness of each of the first three techniques.

We show the predicted redshift distributions for sGRBs in Fig.\ \ref{f:grb_weighted} for each of the three time-delay models considered ($P(\Delta t)\propto (\Delta t)^{-1.5}$, $(\Delta t)^{-1}$, and $(\Delta t)^{-0.5}$).  All three models are in excellent agreement with the redshift distribution from \textit{Swift} in \cite{Fong13}; Kolmogorov-Smirnov (K-S) tests show no significant discrepancies ($p>0.48$ in all cases) between the data and models.  This may be surprising, considering the significant differences between models present in Fig.\ \ref{f:grb_rate}.  However, as the bottom panel of Fig.\ \ref{f:grb_weighted} shows, \textit{Swift}'s sensitivity to sGRBs occurring at $z>0.5$ is very poor.  In addition, the volume of the universe at $z<0.25$ is not large enough for many sGRBs to occur.  The effective redshift leverage is therefore too small for the differences in Fig.\ \ref{f:grb_rate} to appear without obtaining redshifts for many more sGRBs.  We note also that the modeled redshift distribution is sensitive to the assumed intrinsic sGRB luminosity function (Appendix \ref{a:sensitivity}), which is an additional source of uncertainty with this approach.

We have also calculated expected stellar masses for sGRB host galaxies as a function of the time-delay model (Fig.\ \ref{f:grb_smfs}, top panel).  The stellar mass functions show a clear dependence on the time delay distribution; the longer the time delay between the initial star formation and the sGRB, the longer the host galaxy has time to grow.  However, the differences are within the typical $\pm 0.3$ dex systematic biases in recovered stellar masses \citep{Conroy09,Behroozi10}.  It is also possible to predict observed galaxy luminosities for each of the models; Fig.\ \ref{f:grb_smfs} shows the predicted luminosities for sGRB host galaxies in the 2MASS Ks band, compared to \cite{Leibler10}.\footnote{\label{ft:leibler}We do not compare to the stellar masses in \cite{Leibler10}, since their methodology for calculating stellar masses does not include variable dust, variable metallicity, or non-instantaneous stellar population histories; these are all important ingredients in recovering stellar masses at even the 0.3 dex level \citep{Conroy09,Pforr12}.}  However, K-S tests do not reveal any discrepancy between the data in \cite{Leibler10} and the predicted luminosities for any of the three time delay models ($p>0.46$ in all cases).

We finally consider sGRB host galaxy star formation.  The main issue with calculating the fractions of early-type (passive) and late-type (star-forming) sGRB host galaxies is that the separate star formation histories of each galaxy type are very uncertain.  It is exceptionally difficult to constrain these histories from individual galaxy observations, because stellar populations older than 1Gyr tend to have very similar colors \citep{Tojeiro07,Conroy09}.  In addition, semi-analytical models have been notoriously unable to reproduce basic galaxy evolution, including the evolution of galaxy stellar mass functions \citep{Lu12,YLu13,Mutch13}.  However, methods such as comparing stellar mass functions across redshifts \citep{BehrooziCumulative,Salmon13} or more advanced semi-empirical approaches (Behroozi et al., in prep.) may be able to offer better constraints in the future.

That said, we can still provide useful limits.  We assume that early-type and late-type galaxies have the same star formation histories, with the exception of an instantaneous drop in the star formation rate at the redshift of observation for early-type galaxies.  This is guaranteed to overestimate the fraction of early-type hosts, since early-type galaxies have older stellar populations and all the time-delay models predict a decaying sGRB rate with increasing stellar population age.  We can then calculate the early-type vs.\ late-type fraction using the redshift-dependent fraction of early-type vs. late-type galaxies as a function of stellar mass and redshift, from \cite{Brammer11}.  The resulting predictions for sGRB host galaxy specific star formation rates are shown in Fig.\ \ref{f:grb_ssfrfs}, compared to the early/late-type fractions for sGRB host galaxies from \cite{Fong13}.

As with the previous methods, the time delay models are not distinguishable with present data.  In contrast with other methods, the error bars and limits can be improved without requiring any more sGRB statistics.  The primary error bars in the data \citep{Fong13} come from a handful of unclassified host galaxies, which could be classified with deeper follow-up observations.  With very modest constraints on the recent star formation history of elliptical galaxies (e.g., that their stellar ages are $>$1 Gyr older than late-type galaxies), the available sGRB host data could rule out the shortest time-delay model ($P(\Delta t) \propto (\Delta t)^{1.5}$) with $>90\%$ confidence.

\subsection{Local Stellar Mass Function Predictions and Implications for Gravitational Wave Detection Followup}
\label{s:smf_predictions}
\label{s:gravitational_waves}

In Fig.\ \ref{f:grb_smf}, we show predicted probability distributions for sGRB host galaxy stellar masses (i.e., stellar mass functions) in the nearby universe ($z<0.2$).  As with the stellar mass functions in \S \ref{s:time_delay}, longer time delay distributions result in larger host galaxies.  The host stellar mass functions also show some dependence on the progenitor kick distribution, since higher kick velocities preferentially eject sGRB progenitors from low stellar-mass galaxies.  However, these differences are much smaller than for present uncertainties on the time-delay distribution (\S \ref{s:time_delay}).

If gravitational waves are registered by a ground-based detector, such as Advanced LIGO, it will be important to classify the source of the waves by optical and other follow-up observations.  To exclude unrelated astrophysical transients from follow-up, several papers have suggested restricting follow-up to sources close in the sky to a specific catalog of nearby galaxies \citep[e.g.,][]{Metzger13,Hanna13}.  If sGRBs or other star formation-correlated events are strong gravitational wave sources, Fig.\ \ref{f:grb_smf} suggests that the follow-up should be \textit{weighted} towards more likely source galaxies to improve the chance of rapid follow-up.  For example, Fig.\ \ref{f:grb_smf} implies that the vast majority of local sGRBs occur in galaxies larger than $10^{9.5}\Msun$, regardless of model.  Hence, transients occurring near larger galaxies should be prioritized for follow-up observations.  Because galaxy number density increases with decreasing stellar mass, observing for transients near smaller galaxies will result in many fewer successful follow-ups for the same amount of observing time.  In addition, sGRB progenitors formed in $<10^{9.5}\Msun$ galaxies are likely to be unbound from the galaxy/halo potential well (Fig.\ \ref{f:grb_ef}), so the bursts themselves could appear up to a Mpc away \citep{Kelley10}.  Hence, follow-up observations weighted towards more-probable galaxy sources will locate the true source much more quickly than an unweighted galaxy position mask, although the amount of improvement depends on the follow-up telescopes' fields of view and the existence of a complete galaxy catalog \citep{Hanna13}.

\subsection{The Distribution of Compact Binary Mergers Assembled in Globular Clusters}

\label{s:globular}

Globular clusters (GCs) may represent an important source of neutron star--black-hole mergers \citep{Sadowski08}.  Depending on uncertainties in the kick velocity distribution for neutron stars, they may even contribute a significant fraction of neutron star binary mergers as well.  Since it is very difficult to observe globular clusters beyond the local universe \citep{Harris13}---and hence, to estimate their distribution in the potential well of the host halo---we cannot quantitatively predict the locations of sGRBs generated in globular clusters.  That said, several qualitative predictions are possible.

Not all sGRBs can be formed in GC systems, because the half-light radii of GC systems are typically five times larger than the effective radii of their host galaxies \citep{Kartha14}.  At face value, the \cite{Fong13b} results would put an upper limit of $\approx40\%$ for the fraction of sGRBs which can form in GCs; as noted in \cite{Salvaterra10}, the interstellar medium in GCs is dense enough to make afterglows which would be observable by \textit{Swift}.  If indeed most sGRBs with large offsets are generated in GCs, there would be several effects on the host galaxy demographics.  For example, the redshift distribution of GC-associated sGRBs could be very different from galaxy-associated sGRBs, as the dynamical formation of compact binaries in GCs would presumably not track the cosmic star formation history.  In addition, the offsets (relative to host galaxy effective radius) for GC-associated sGRBs would not correlate with host galaxy mass; however, if sGRBs at large offsets are due instead to high kick velocities, then Fig.\ \ref{f:grb_ef} shows that there should be a significant anticorrelation between galaxy mass and sGRB offset.

Since there are only four sGRBs with offsets securely measured at $>5\Reff$ \citep{Fong13b}, none of these statistical tests are currently possible.\footnote{We note that \cite{Fong13b} do attempt a correlation test between galaxy mass and sGRB offset, but they use the stellar masses from \cite{Leibler10}; as discussed in footnote \ref{ft:leibler}, these stellar masses may have significant additional scatter with respect to the true galaxy stellar masses.}  However, future data should provide a means to separate between GC-associated and galaxy-associated origins for high-offset sGRBs.

\section{Conclusions}

\label{s:conclusions}

We have presented a method for robustly modeling short gamma-ray bursts (sGRBs) in the context of $\Lambda$CDM cosmology and hierarchical halo growth.  Key findings include:
\begin{enumerate}
\item The observed hostless fraction of $\gtrsim$ 18\% \citep{Fong13b} means that white dwarf/white dwarf mergers are strongly disfavored as sGRB progenitors (\S \ref{s:progenitors}).
\item The best-fitting match to the observed sGRB --- galaxy offset distribution in \citep{Fong13b} would require 19\% of sGRB progenitors to receive natal kicks of more than 150 km s$^{-1}$ (\S \ref{s:progenitors}).
\item The expected hostless fraction is robust to a large number of uncertainties and systematics, including the assumed time-delay distribution (\S \ref{s:hostless_results}, Table \ref{t:hostless_fracs}), the \textit{Swift} redshift sensitivity, and observational follow-up of host galaxies (\S \ref{s:systematics}, Table \ref{t:hostless_fracs_syst}).
\item Regardless of the assumed time-delay distribution, most sGRBs occur in a tight halo mass range around $10^{12}\Msun$, similar to the halo mass range where most stars are formed (\S \ref{s:halo_results}, Fig.\ \ref{f:grb_halos}).
\item Most sGRBs occur in a tight stellar mass range around 5$\times 10^{10.5}\Msun$), with only a weak dependence on the time-delay model (\S \ref{s:time_delay}).
\item The observed redshift distribution of sGRBs also does not offer strong constraints on the time-delay model (\S \ref{s:time_delay}), due to the limited redshift sensitivity of \textit{Swift}.
\item The clearest observable impact of the time-delay model may be on the specific star formation rates of sGRB hosts (\S \ref{s:time_delay}).
\item The observed redshift distribution of hostless sGRBs is extremely similar to the observed redshift distribution of sGRBs with galaxy hosts for $z<3$, regardless of model (Fig.\ \ref{f:grb_ef_models}).
\item The vast majority of sGRBs are expected to occur in galaxies with stellar mass $>10^{9.5}\Msun$, which suggests that optical/EM follow-up of gravitational wave sources may be made more efficient by primarily searching near such galaxies (\S \ref{s:gravitational_waves}).
\end{enumerate}

\acknowledgments
Support for PSB was provided in part by an HST Theory grant; program number HST-AR-12159.01-A was provided by NASA through a grant from the Space Telescope Science Institute, which is operated by the Association of Universities for Research in Astronomy, Incorporated, under NASA contract NAS5-26555.  PSB was also supported by a Giacconi Fellowship through the Space Telescope Science Institute.  ERR acknowledges support from the David and Lucile Packard Foundation and  NSF grant: AST-0847563. CLF was supported through the auspices of the National Nuclear Security Administration of the U.S. Department of Energy, and supported by its contract DEAC52-06NA25396 at Los Alamos National Laboratory.  We appreciate the insightful comments we have received from the anonymous referee as well as Edo Berger, Andy Fruchter, Nick Gnedin, Dan Holz, Luke Kelley, and Andrey Kravtsov during the preparation of this paper.  ERR's views on the topic have been clarified through many discussions with Ilya Mandel.\\

\bibliography{master_bib}

\appendix
\begin{figure*}
\begin{center}
\plotminigrace{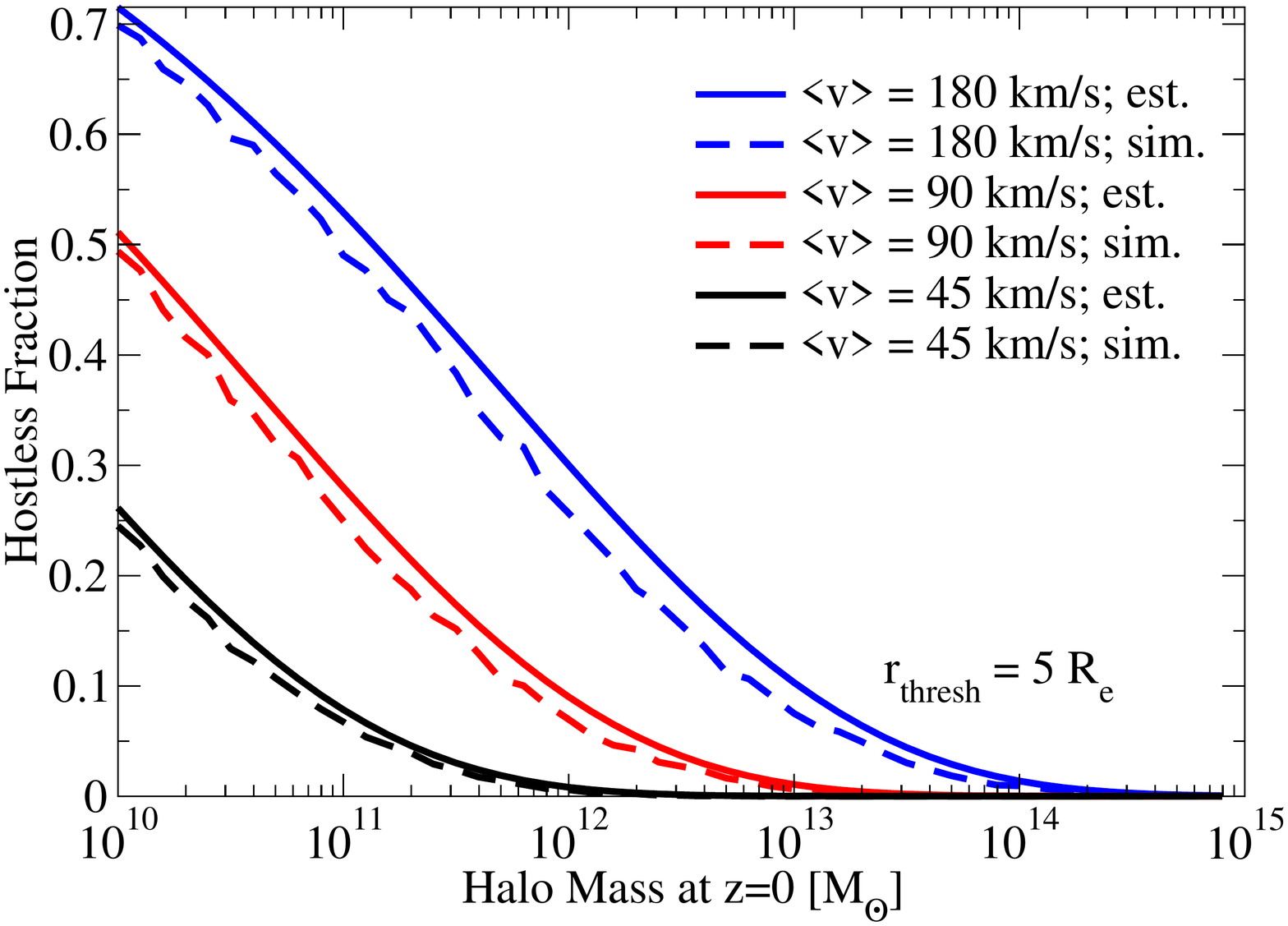}\plotminigrace{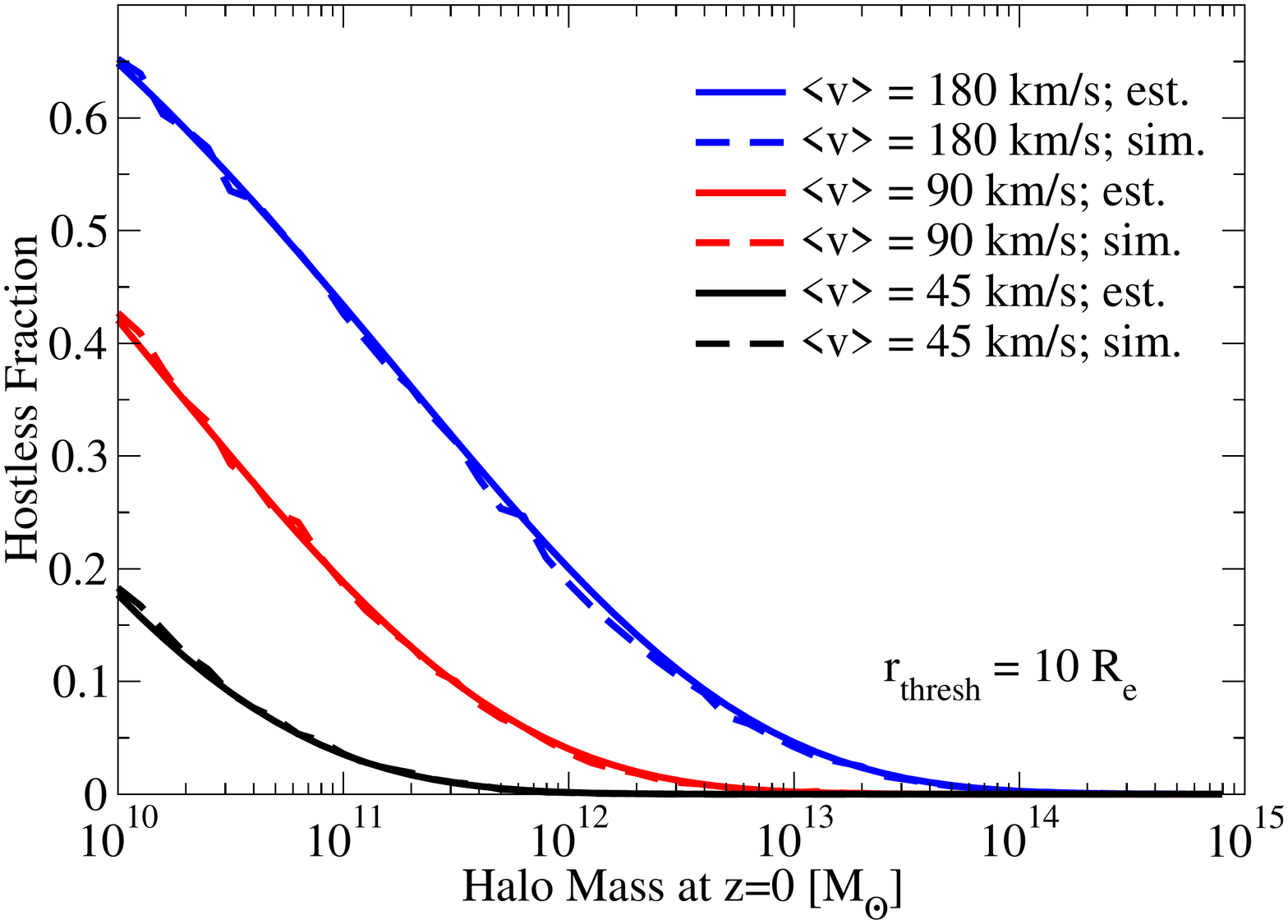}\\
\plotminigrace{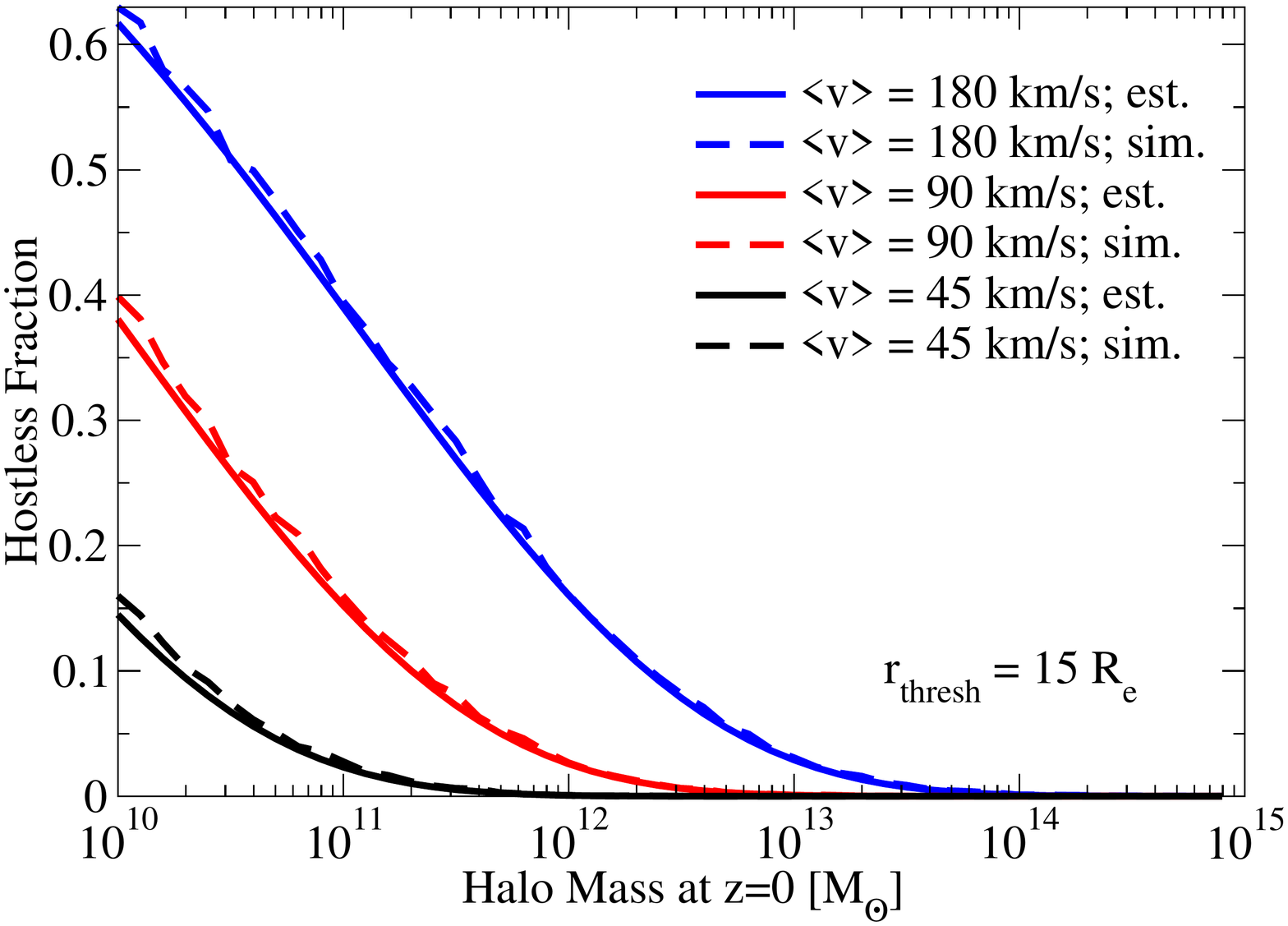}
\end{center}
\caption{Simulated sGRB hostless fractions (``\textbf{sim.}'') from Appendix \ref{s:tests} compared to the estimated fractions (``\textbf{est.}'') from \S \ref{s:vel_methods}, obtained for progenitor velocity kicks occurring at $z=1$ and sGRB detonation at $z=0$.  $\langle v \rangle$ refers to the average speed used in the exponential velocity kick distribution; $r_\mathrm{thresh}$ is the projected radius at which a sGRB would be considered hostless (see \S \ref{s:vel_methods}).}
\label{f:sim_vs_est}
\end{figure*}

\section{Derivation of $SFR(M_h, z)$}
\label{s:sfr}

We defer to \cite{BWC13} (hereafter, BWC13) for the full derivation of $SFR(M_h,z)$, but we provide a summary of the technique here.  BWC13 adopts a very flexible parametrization for the stellar mass -- halo mass -- redshift relation, $SM(M_h, z)$.\footnote{The relation at $z=0$ is controlled by six parameters.  These include characteristic stellar and halo masses, a faint-end slope, a massive-end cutoff, a transition region shape, and scatter in stellar mass at fixed halo mass.  For each of these six parameters, another two parameters control the evolution to intermediate ($z \approx 1$) and high ($z > 3$) redshift.}  A particular point in the $SM(M_h, z)$ function parameter space implies an assignment of galaxy stellar masses to all halos at all redshifts for a dark matter simulation; our simulation and halo catalogs are described in \cite{Bolshoi} and \cite{Rockstar,BehrooziTree}.  From these assigned stellar masses, predictions for the stellar mass function (i.e., the number density of galaxies as a function of stellar mass) can be made; in addition, growth in the assigned stellar masses along dark matter halo merger trees can be used to predict galaxy star formation rates.  BWC13 compares these predicted observables to a wide range of published data from $z=0$ to $z=8$ and uses a Markov Chain Monte Carlo (MCMC) algorithm to derive the posterior distribution for $SM(M_h, z)$ (and the resulting implied $SFR(M_h,z)$).  The resulting best-fit $SFR(M_h,z)$ is consistent with all recent published stellar mass functions, specific star formation rates, and cosmic star formation rates over this entire redshift range; in addition, the best-fit stellar mass---halo mass relation, $SM(M_h,z)$, is consistent with a wide range of independent techniques (see BWC13).

\section{Tests of the Gravitational Potential Method}
\label{s:tests}

We present tests of the model in \S \ref{s:vel_methods} with a set of more realistic particle simulations.  These simulations use the same evolving spherical halo potential as in \cite{BehrooziUnbound}, and we refer readers to Appendix B of that paper for full details.  To summarize the technique, we inject sGRB progenitors into spherical NFW halos at $z=1$ (mass range: $10^{10}\Msun < M_h < 10^{14}\Msun$) at a radius of $0.015 R_{200c}$ with the circular velocity of the halo at that location.  We then apply a velocity kick drawn from an exponential distribution with mean speed $\langle v\rangle$ of 45, 90, or 180 km s$^{-1}$, and simulate the trajectory of the progenitor to $z=0$ using leapfrog integration.  We evolve the halo mass during this time according to the median mass accretion histories in \cite{BWC13}, and we use a force law which accounts for the small effect of Hubble expansion on top of the Newtonian force from the halo \citep{BehrooziUnbound}.  Finally, we compute the projected radius at $z=0$ for the sGRB as viewed along a random line of sight.  We repeat this process for 10,000 realizations for each halo mass, velocity distribution, and threshold radius.

We find that the upper limit estimations from \S \ref{s:vel_methods} are extremely close to the true simulated hostless fractions, as shown in Fig.\ \ref{f:sim_vs_est}.  The proximity of the estimates to the simulations is due to several factors.  First, particles in gravitational orbits tend to spend most of their time near the turnaround radius, meaning that comparing the turnaround radius to the hostless threshold radius (as in \S \ref{s:vel_methods}) is a fairly accurate approximation.  Secondly, because sGRBs which begin at $\Reff$ and reach at least $5\Reff$ must be on fairly radial trajectories, the assumption that all of the kick energy is directed radially in \S \ref{s:vel_methods} is also very good one.  Finally, the fact that \S \ref{s:vel_methods} considered the 3D turnaround radius instead of the 2D projected radius offers a partial correction for the effects of Hubble expansion over the period from $z=1$ to $z=0$.  Specifically, the accelerated expansion of the universe over this time period results in a force law which is weakened relative to the Newtonian version.  Close to the halo center, this effect is small, but it becomes more important at larger distances \citep{BehrooziUnbound}; this explains why the simulations show a slight excess over the estimated upper limits for $r_\mathrm{thresh}=15 \Reff$, but are still well within the estimates for $r_\mathrm{thresh}=5\Reff$.  The chosen time period ($z=1$ to $z=0$) maximizes the severity of this effect, because the expansion of the universe is accelerating over this entire time period.  At redshifts $z>1$, the density of dark energy is small enough that the universe is still decelerating, which makes the force law \textit{stronger} than the Newtonian version; this would make it \textit{harder} for progenitors to escape from the original galaxy.

\section{Data for Quantitative Effects of Systematics on Hostless Fractions}
\label{s:quant}

See tabulated results in Table \ref{t:hostless_fracs_syst}.  The ``True'' column indicates the mean hostless fraction of all sGRBs from $z=\infty$ to $z=0$ occurring in a fixed comoving volume of the universe.  The ``Perfect Obs.'' column gives the hostless fraction for perfect observations; this includes the effects of increasing visible comoving volumes at higher redshift as well as time dilation, but it assumes a perfect sGRB detector and follow-up observations to arbitrarily faint host galaxy luminosities.  The ``True for \textit{Swift} Obs.'' column modifies the ``Perfect Obs.'' column to include the redshift sensitivity for the \textit{Swift} satellite from Eq.\ \ref{e:detector}, but again assumes perfect follow-up observations.  Other columns are discussed in \S \ref{s:systematics}.

\begin{table}
\begin{center}
\caption{sGRB Hostless Fractions, Sensitivity to Systematics}
\label{t:hostless_fracs_syst}
\begin{tabular}{cccccccc}
\hline
\hline
$\langle v \rangle$ km s$^{-1}$ & $r_{thresh}$ ($\Reff$) & n & True & Perfect Obs. & True for Swift Obs. & \textbf{Actual Obs.} & Upgraded Swift\\
\hline
45 & 15 & -0.5 & <0.011 & <0.010 & <0.012 & <0.036 & <0.054\\
45 & 15 & -1.0 & <0.013 & <0.012 & <0.015 & <0.049 & <0.068\\
45 & 15 & -1.5 & <0.015 & <0.014 & <0.019 & <0.061 & <0.077\\
45 & 10 & -0.5 & <0.015 & <0.015 & <0.016 & <0.039 & <0.057\\
45 & 10 & -1.0 & <0.018 & <0.018 & <0.020 & <0.053 & <0.070\\
45 & 10 & -1.5 & <0.021 & <0.020 & <0.025 & <0.065 & <0.080\\
45 & 5 & -0.5 & <0.030 & <0.033 & <0.030 & <0.050 & <0.067\\
45 & 5 & -1.0 & <0.037 & <0.039 & <0.039 & <0.067 & <0.082\\
45 & 5 & -1.5 & <0.042 & <0.043 & <0.047 & <0.082 & <0.093\\
\hline
90 & 15 & -0.5 & <0.052 & <0.053 & <0.053 & <0.070 & <0.083\\
90 & 15 & -1.0 & <0.062 & <0.060 & <0.067 & <0.091 & <0.100\\
90 & 15 & -1.5 & <0.069 & <0.066 & <0.079 & <0.110 & <0.112\\
90 & 10 & -0.5 & <0.069 & <0.073 & <0.070 & <0.085 & <0.099\\
90 & 10 & -1.0 & <0.082 & <0.083 & <0.087 & <0.109 & <0.117\\
90 & 10 & -1.5 & <0.092 & <0.090 & <0.102 & <0.130 & <0.130\\
90 & 5 & -0.5 & <0.121 & <0.134 & <0.120 & <0.133 & <0.148\\
90 & 5 & -1.0 & <0.143 & <0.150 & <0.145 & <0.164 & <0.172\\
90 & 5 & -1.5 & <0.158 & <0.161 & <0.168 & <0.191 & <0.188\\
\hline
180 & 15 & -0.5 & <0.176 & <0.184 & <0.177 & <0.187 & <0.196\\
180 & 15 & -1.0 & <0.201 & <0.202 & <0.209 & <0.224 & <0.222\\
180 & 15 & -1.5 & <0.219 & <0.214 & <0.237 & <0.255 & <0.241\\
180 & 10 & -0.5 & <0.214 & <0.228 & <0.213 & <0.223 & <0.235\\
180 & 10 & -1.0 & <0.244 & <0.249 & <0.249 & <0.262 & <0.263\\
180 & 10 & -1.5 & <0.264 & <0.263 & <0.279 & <0.296 & <0.283\\
180 & 5 & -0.5 & <0.307 & <0.333 & <0.304 & <0.311 & <0.329\\
180 & 5 & -1.0 & <0.344 & <0.358 & <0.345 & <0.355 & <0.361\\
180 & 5 & -1.5 & <0.370 & <0.375 & <0.379 & <0.393 & <0.384\\
\hline
\end{tabular}

\tablecomments{Upper limits on the observed fraction of hostless sGRBs for a range of different assumptions for the mean progenitor kick velocity ($\langle v\rangle$; \S \ref{s:vel_methods}), the hostless threshold radius ($r_\mathrm{thresh}$; \S \ref{s:vel_methods}), and the time-delay power law index ($n$; \S \ref{s:halo_masses}), analogous to Table \ref{t:hostless_fracs}.  The \textbf{True} column gives the fraction of hostless sGRBs in the entire universe.  The \textbf{Perfect Obs.} column gives the hostless fraction for sGRBs detectable by a perfect observatory; this includes both the effect of the increasing comoving volume available at higher redshifts and the time dilation between sGRB events at higher redshifts (Eq.\ \ref{e:sgrbr_obs}).  \textbf{True for Swift Obs.} gives the fraction of \textit{Swift}-detected sGRBs which would be hostless; this adds the \textit{Swift} sGRB detection probability from Eq.\ \ref{e:detector}.  \textbf{Actual Obs.} gives the fraction of \textit{Swift}-detected sGRBs for which follow-up to F160W $<$ 26 mag would not find a nearby host, identical to Table \ref{t:hostless_fracs}.  Finally, \textbf{Upgraded Swift} shows the expected hostless fractions if \textit{Swift} were made dramatically more sensitive to higher-redshift sGRBs without corresponding increases in the depth of follow-up observations (see \S \ref{s:systematics}).}
\end{center}
\end{table}

\begin{figure}
\begin{center}
\plotminigrace{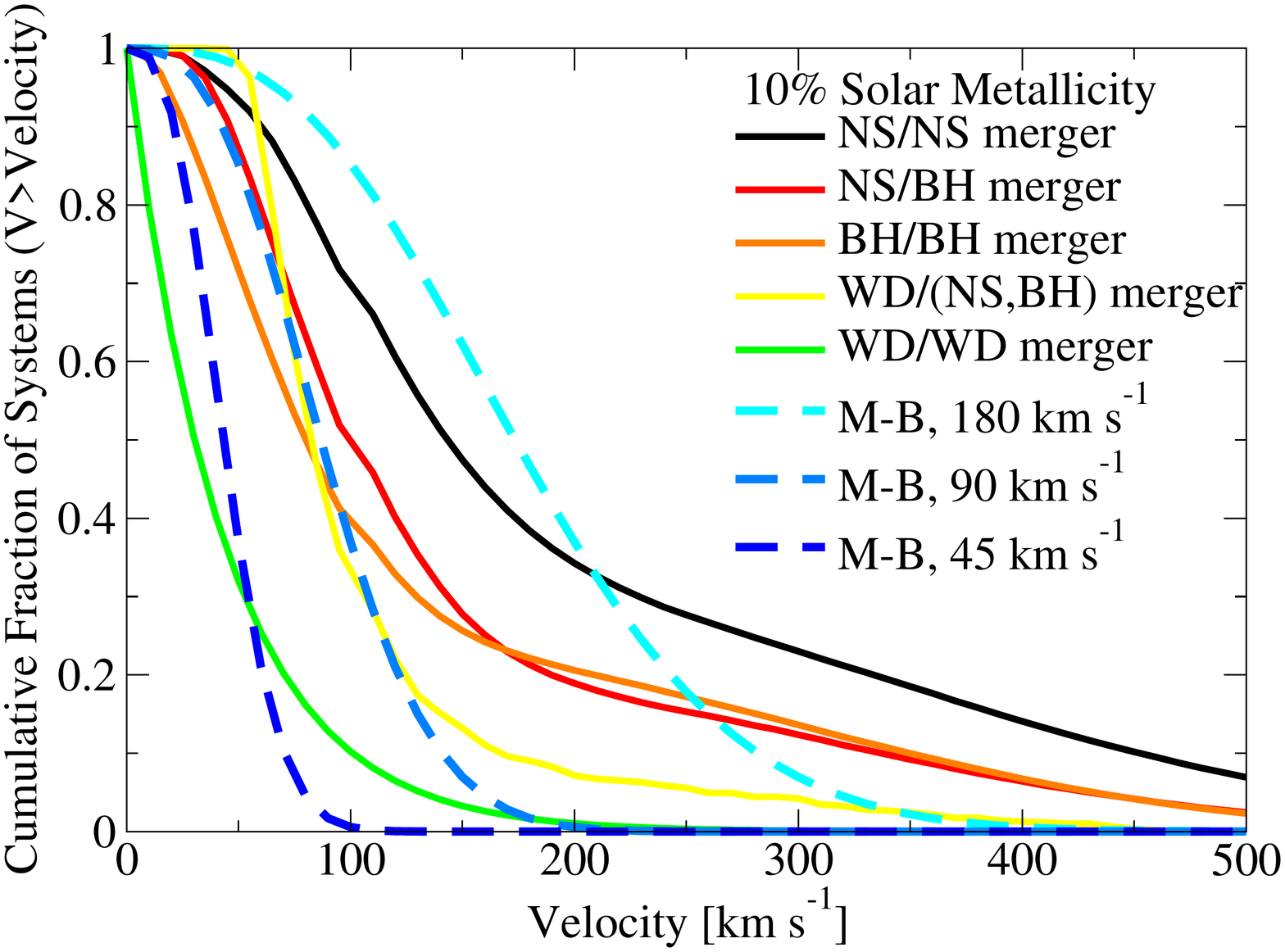}\plotminigrace{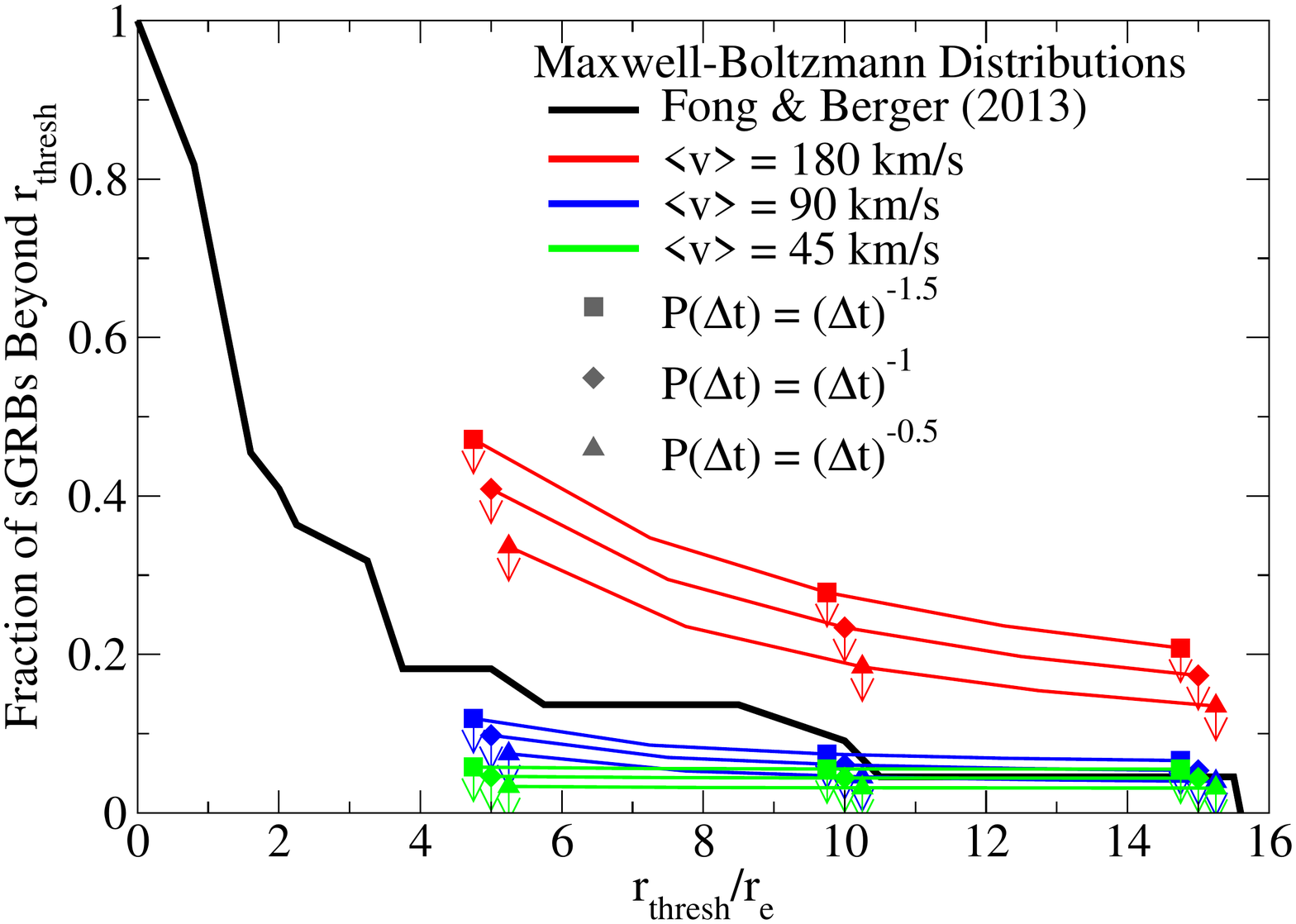}
\end{center}
\caption{\textbf{Left} panel: Analogous to Fig.\ \ref{f:vel_dist}. \textit{Dashed lines} show Maxwell-Boltzmann (MB) distributions for comparison with $\langle v \rangle$ of 180, 90, and 45 km s$^{-1}$.  \textbf{Right} panel: Analogous to Fig.\ \ref{f:fong_comp}, for the Maxwell-Boltzmann velocity distributions shown in the left panel.  The black line shows the observed location distribution of sGRBs from \cite{Fong13b} and the predicted locations of sGRBs from Table \ref{t:hostless_fracs_syst_mb} for several assumptions about the sGRB progenitor kick velocities and sGRB merger time delays.  Compared to Fig.\ \ref{f:fong_comp}, the 90 km s$^{-1}$ MB models are predicted to have a significantly lower fraction of sGRBs beyond $5\Reff$, whereas the 180 km s$^{-1}$ MB models have a much higher fraction of sGRBs at large offsets.  This suggests that the offset distribution of sGRBs is especially sensitive to the fraction of natal kicks above about 150 km s$^{-1}$.}
\label{f:vel_dist_maxwell}
\end{figure}

\begin{table}
\begin{center}
\caption{sGRB Hostless Fractions, Sensitivity to Systematics; Maxwell-Boltzmann Distribution}
\label{t:hostless_fracs_syst_mb}
\begin{tabular}{cccccccc}
\hline
\hline
$\langle v \rangle$ km s$^{-1}$ & $r_{thresh}$ ($\Reff$) & n & True & Perfect Obs. & True for Swift Obs. & \textbf{Actual Obs.} & Upgraded Swift\\
\hline
45 & 15 & -0.5 & <0.006 & <0.004 & <0.007 & <0.031 & <0.052\\
45 & 15 & -1.0 & <0.007 & <0.005 & <0.009 & <0.044 & <0.065\\
45 & 15 & -1.5 & <0.008 & <0.006 & <0.011 & <0.055 & <0.074\\
45 & 10 & -0.5 & <0.008 & <0.006 & <0.009 & <0.032 & <0.052\\
45 & 10 & -1.0 & <0.009 & <0.008 & <0.012 & <0.044 & <0.065\\
45 & 10 & -1.5 & <0.011 & <0.009 & <0.014 & <0.055 & <0.074\\
45 & 5 & -0.5 & <0.014 & <0.014 & <0.015 & <0.033 & <0.052\\
45 & 5 & -1.0 & <0.018 & <0.017 & <0.020 & <0.046 & <0.065\\
45 & 5 & -1.5 & <0.020 & <0.020 & <0.025 & <0.058 & <0.075\\
\hline
90 & 15 & -0.5 & <0.026 & <0.025 & <0.027 & <0.040 & <0.055\\
90 & 15 & -1.0 & <0.031 & <0.029 & <0.036 & <0.053 & <0.068\\
90 & 15 & -1.5 & <0.035 & <0.032 & <0.044 & <0.066 & <0.077\\
90 & 10 & -0.5 & <0.034 & <0.035 & <0.035 & <0.045 & <0.059\\
90 & 10 & -1.0 & <0.041 & <0.042 & <0.046 & <0.060 & <0.072\\
90 & 10 & -1.5 & <0.047 & <0.047 & <0.056 & <0.074 & <0.081\\
90 & 5 & -0.5 & <0.069 & <0.078 & <0.069 & <0.075 & <0.082\\
90 & 5 & -1.0 & <0.085 & <0.093 & <0.090 & <0.098 & <0.099\\
90 & 5 & -1.5 & <0.097 & <0.104 & <0.109 & <0.119 & <0.111\\
\hline
180 & 15 & -0.5 & <0.128 & <0.133 & <0.132 & <0.135 & <0.131\\
180 & 15 & -1.0 & <0.155 & <0.153 & <0.169 & <0.173 & <0.154\\
180 & 15 & -1.5 & <0.174 & <0.167 & <0.202 & <0.208 & <0.171\\
180 & 10 & -0.5 & <0.180 & <0.194 & <0.182 & <0.184 & <0.182\\
180 & 10 & -1.0 & <0.218 & <0.222 & <0.231 & <0.234 & <0.213\\
180 & 10 & -1.5 & <0.246 & <0.242 & <0.274 & <0.278 & <0.236\\
180 & 5 & -0.5 & <0.340 & <0.378 & <0.335 & <0.336 & <0.348\\
180 & 5 & -1.0 & <0.403 & <0.424 & <0.407 & <0.409 & <0.399\\
180 & 5 & -1.5 & <0.449 & <0.455 & <0.470 & <0.471 & <0.435\\
\hline
\end{tabular}
\tablecomments{Analogous to Table \ref{t:hostless_fracs_syst}, except with Maxwell-Boltzmann velocity kick distributions instead of exponential velocity distributions.  Upper limits on the observed fraction of hostless sGRBs for a range of different assumptions for the mean progenitor kick velocity ($\langle v\rangle$; \S \ref{s:vel_methods}), the hostless threshold radius ($r_\mathrm{thresh}$; \S \ref{s:vel_methods}), and the time-delay power law index ($n$; \S \ref{s:halo_masses}), analogous to Table \ref{t:hostless_fracs}.  The \textbf{True} column gives the fraction of hostless sGRBs in the entire universe.  The \textbf{Perfect Obs.} column gives the hostless fraction for sGRBs detectable by a perfect observatory; this includes both the effect of the increasing comoving volume available at higher redshifts and the time dilation between sGRB events at higher redshifts (Eq.\ \ref{e:sgrbr_obs}).  \textbf{True for Swift Obs.} gives the fraction of \textit{Swift}-detected sGRBs which would be hostless; this adds the \textit{Swift} sGRB detection probability from Eq.\ \ref{e:detector}.  \textbf{Actual Obs.} gives the fraction of \textit{Swift}-detected sGRBs for which follow-up to F160W $<$ 26 mag would not find a nearby host, identical to Table \ref{t:hostless_fracs}.  Finally, \textbf{Upgraded Swift} shows the expected hostless fractions if \textit{Swift} were made dramatically more sensitive to higher-redshift sGRBs without corresponding increases in the depth of follow-up observations (see \S \ref{s:systematics}).}
\end{center}
\end{table}

\section{Results for Maxwell-Boltzmann Velocity Distributions}
\label{a:maxwell}

To test how the velocity distribution's function form affects our conclusions, we have repeated our full analysis using Maxwell-Boltzmann velocity kick distributions instead of exponential distributions.  As shown in Fig.\ \ref{f:vel_dist_maxwell}, Maxwell-Boltzmann velocity distributions do not have the high-velocity tails present in the exponential distribution, and so represent a poorer match to the theoretical models.  Table \ref{t:hostless_fracs_syst_mb} lists the expected hostless fractions for all models, using Maxwell-Boltzmann distributions with identical mean velocities (45, 90, and 180 km s$^{-1}$) as the exponential distributions in our main analysis.  The main findings for the exponential distributions still hold, including the relative insensitivity of the hostless fraction to the time delay model and any redshift reweighting from geometrical and instrumental effects.  The principal change with the Maxwell-Boltzmann distribution is in the radial distribution of the sGRBs.  Velocities of 45 and 90 km s$^{-1}$ are low with respect to the circular velocities of halos where most star formation occurs ($M_h\approx 10^{12}\Msun$; $v\approx 250$ km s$^{-1}$), and the lack of high-velocity tails mean that extremely few sGRBs would be able to escape their host galaxies.  For the 180 km s$^{-1}$ model, the mean velocity is larger relative to star-forming halos' circular velocity, so many more sGRBs make it beyond 5$\Reff$; however, the lack of high-velocity tails implies that comparatively few sGRBs will appear beyond 10 or 15$\Reff$.  The fact that star formation in halos occurs in a narrow mass range means that there is a correspondingly narrow range of kick energies which will allow sGRB progenitors to escape to a given distance from the host galaxy.  For this reason, the relative fraction of sGRBs at different distances from the host galaxy is a potentially very sensitive probe of the shape of the velocity kick distribution.

\begin{figure}
\plotminigrace{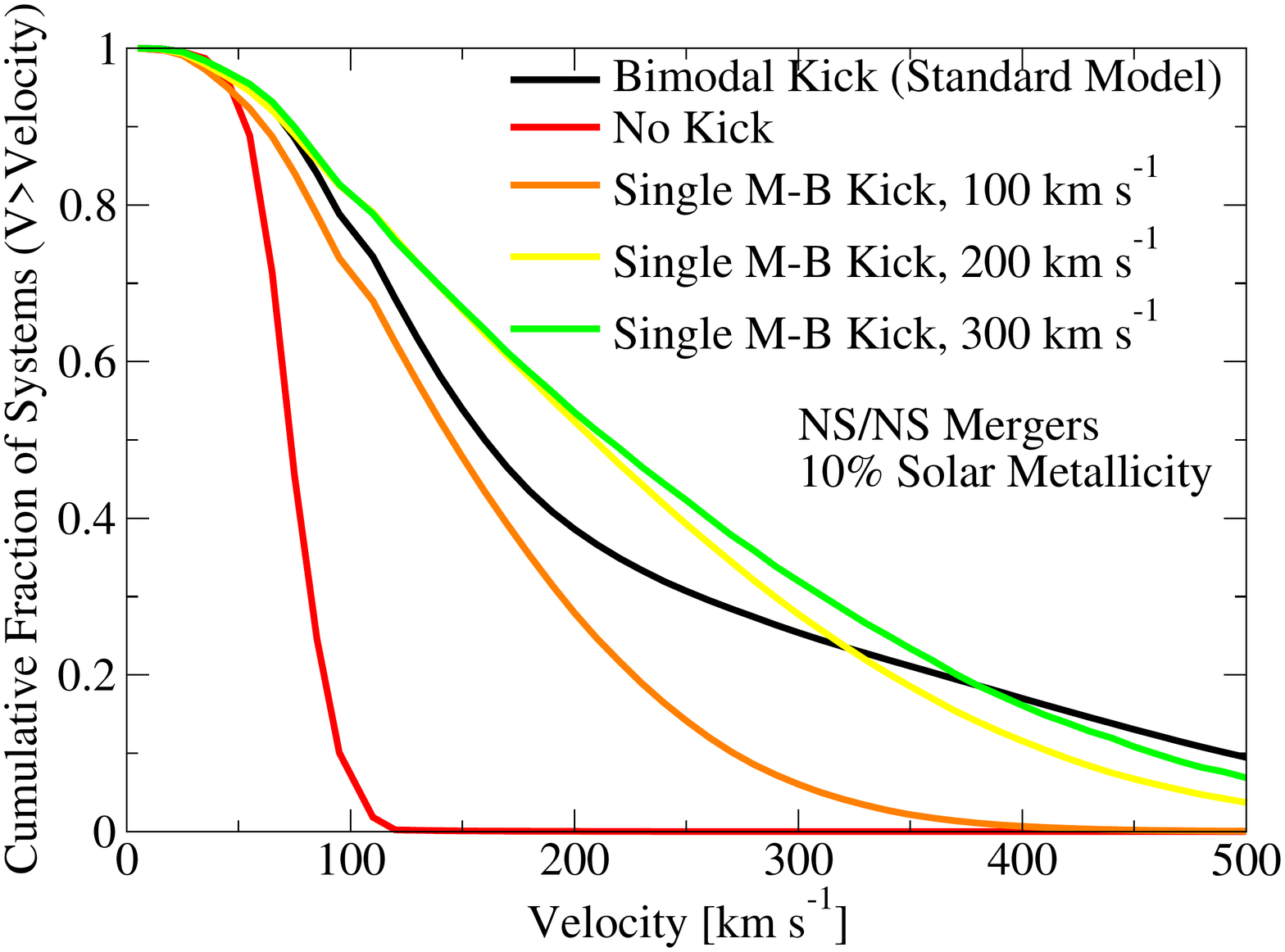}\plotminigrace{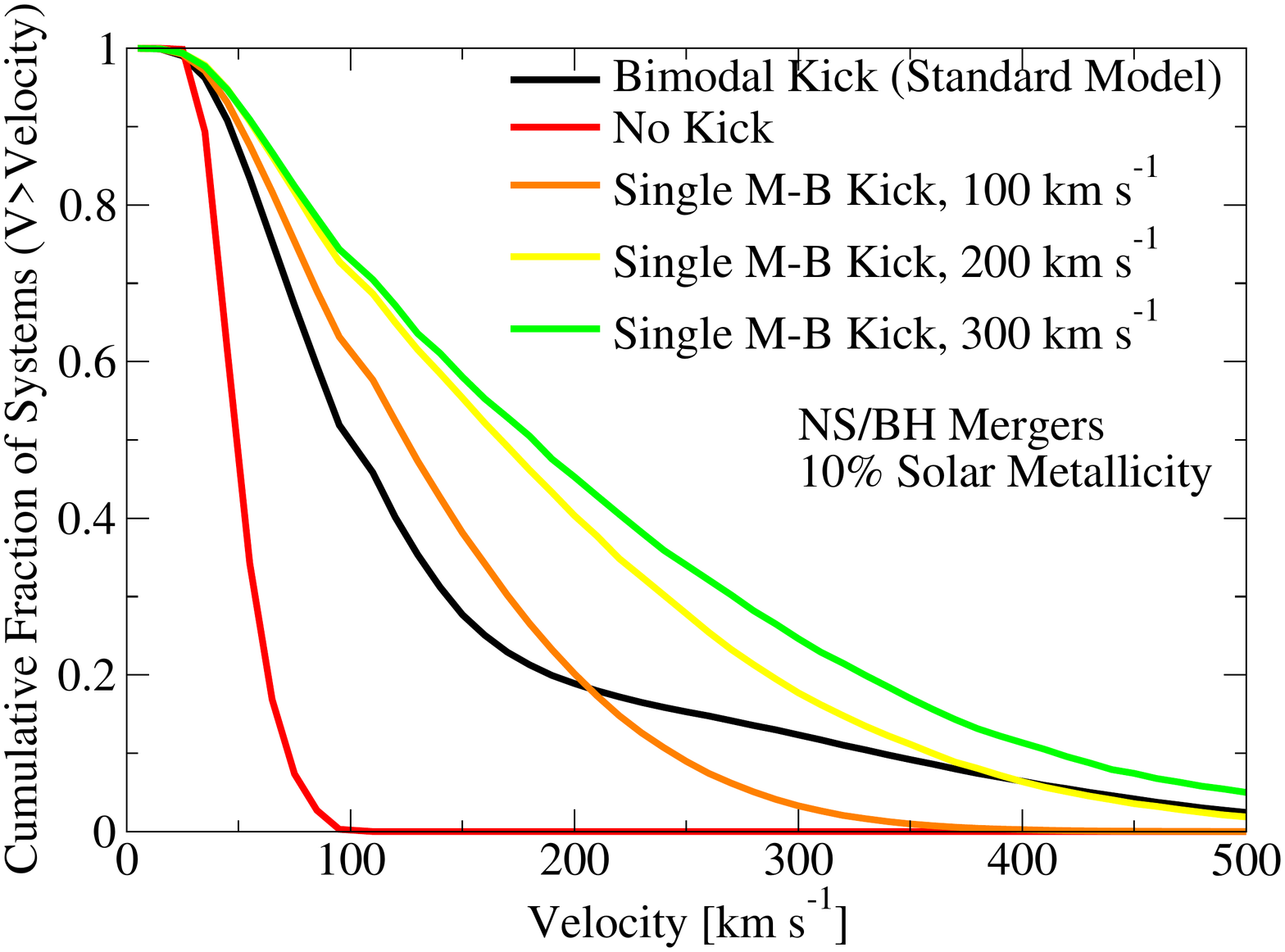}
\caption{\textbf{Left} panel: Cumulative velocity probability distributions for neutron star/neutron star mergers for a range of supernova kick distributions.  These include our standard bimodal kick distribution (see text), no supernova kick, and three single mode Maxwell-Boltzmann kick distributions.  \textbf{Right} panel: Same, for neutron star/black hole mergers.}
\label{f:ns_uncertainties}
\end{figure}

\section{Neutron Star and Black Hole Kick Velocity Uncertainties}
\label{a:ns_uncertainties}

Soon after the discovery of pulsars, it was realized that their velocity distribution was much higher than the normal stellar population \citep{Gunn70}. Models for these enhanced velocities included both asymmetries in the supernova explosion that produced the pulsar \citep{Herant95,Burrows96,Fryer04,Janka07,Endeve13}, neutrino asymmetries \citep{Kusenko04,Socrates05}, and the momentum exchange in unbinding a massive binary during the explosion \citep{Hills83}.  A growing list of compact binary systems seem to require at least some kick in addition to the momentum imparted in a binary \citep{Flannery75,Johnston92,Yamaoka93,Kaspi94,Brandt95,Kaspi96,Bildsten97,Fryer97} and most work assumes that either ejecta or neutrino asymmetries in the supernova impart a kick on the forming neutron star that produces the observed high pulsar space velocities. 

These pulsar velocities can then be used to study the kick produced in a supernova explosion.   \cite{Lyne82} argued that the RMS space velocity of observed pulsars was 210 km s$^{-1}$.  But improved measurements of pulsar distances \citep{Taylor93} along with an increased sample of pulsars led to much larger pulsar RMS velocity estimates of 450 km s$^{-1}$ \citep{Lyne94}.  Selection biases complicate these estimates \citep{Hansen97,Fryer98}, but detailed studies of these effects suggest that the larger RMS velocities are more accurate \citep{Arzoumanian02}.  Such high velocities rule out binary disruption as the primary kick mechanism and place stringent constraints on any explosion asymmetry mechanism.

It is also possible to match observations with a bimodal pulsar velocity distribution; e.g., half the pulsars receiving a Gaussian $\sigma \sim 90$ km s$^{-1}$ kick and the other half receiving a $\sigma \sim$ 500 km s$^{-1}$ kick \citep{Arzoumanian02}.  This bimodal kick also appears to better explain a wide range of neutron star populations and binary systems \citep{Fryer98,Podsiadlowski05}.  We use this bimodal kick distribution as our standard calculation for this study.  However, it is possible that a single Maxwellian can explain the data, e.g., a Gaussian with $\sigma \sim$ 290 km s$^{-1}$ \citep{Arzoumanian02}.  We have conducted a series of single-mode calculations using a range of RMS supernova kick velocities (Fig.\ \ref{f:ns_uncertainties}).  For low kick velocities, momentum conservation following mass loss in the supernova explosion determines the mean motion of the binary systems.  But beyond an RMS velocity of 50--100 km s$^{-1}$, the kick dominates.  The velocity distributions of both neutron star/neutron star and neutron star/black hole mergers are shown in Fig.\ \ref{f:ns_uncertainties}.  Although a ``no kick'' model cannot explain the distribution of sGRBs, RMS kick velocities above $\sim 100$ \rm km s$^{-1}$ can explain the data.  High velocity single-kick models may produce many more ejected sGRBs than observed; however, uncertain afterglow likelihoods for ejected sGRBs (\S \ref{s:progenitors}) substantially weaken constraints on the maximum allowable kick velocities.

\begin{figure}
\plotminigrace{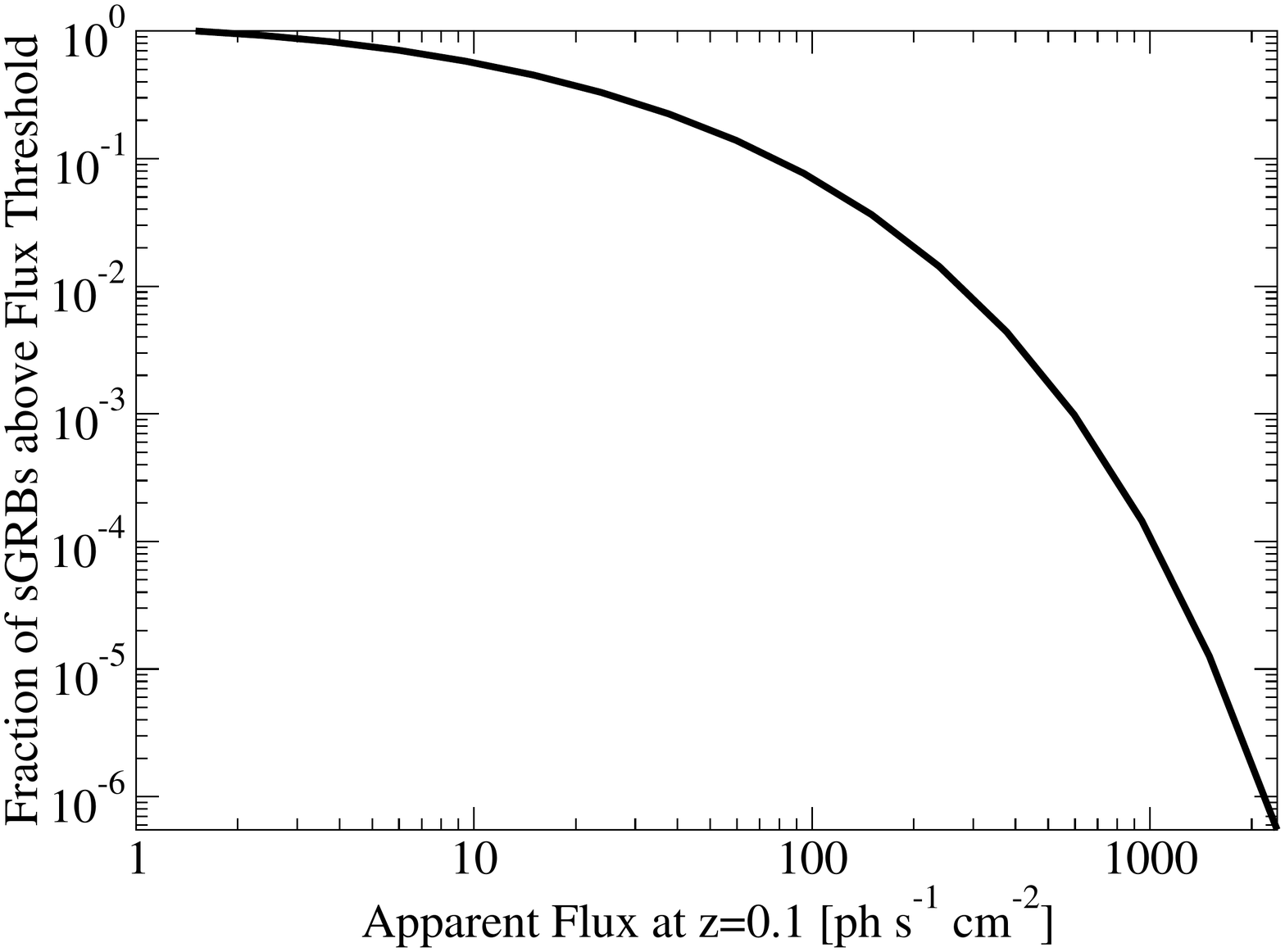}\plotminigrace{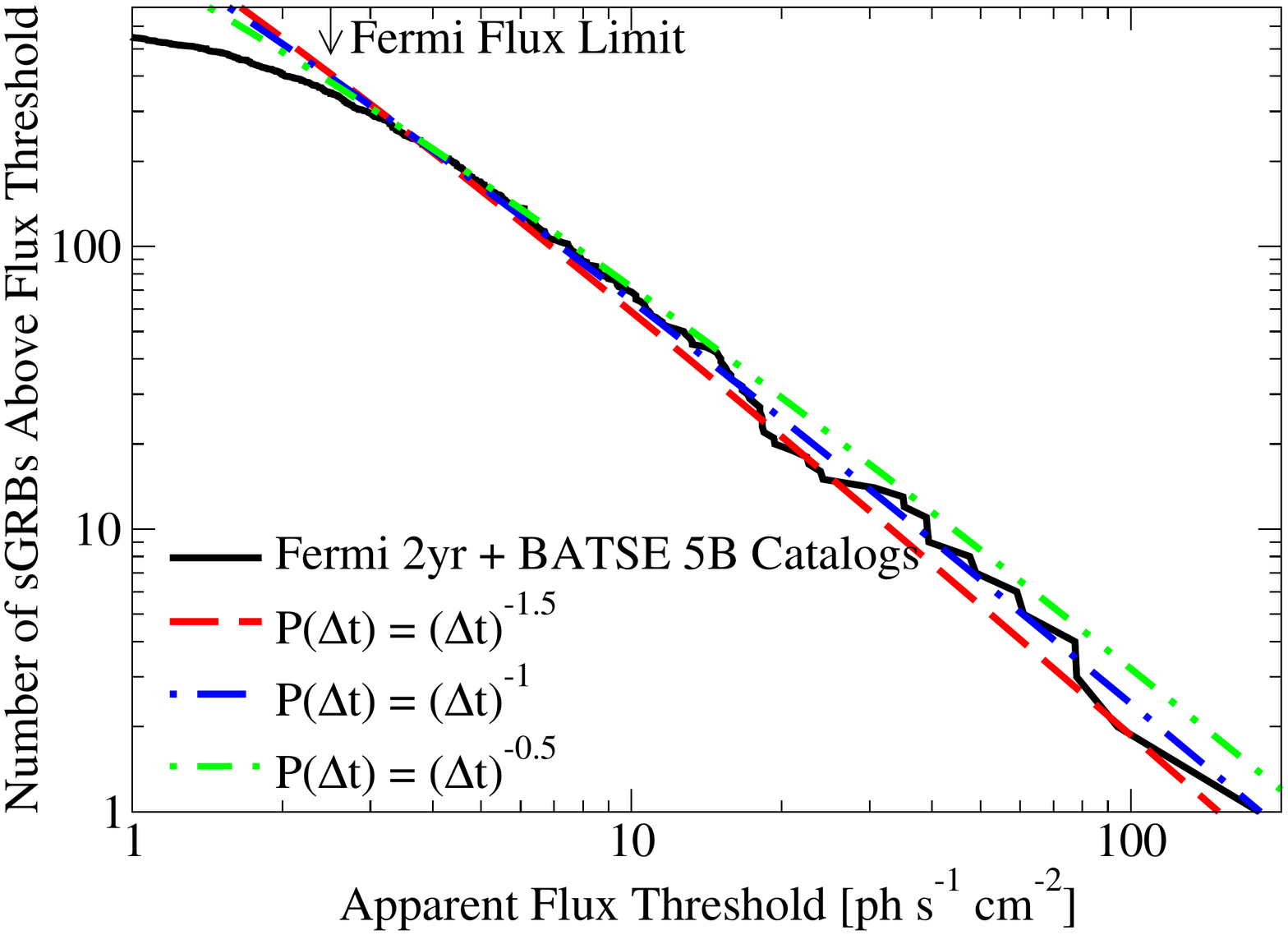}
\caption{\textbf{Left} panel: inferred intrinsic cumulative peak flux function for sGRBs, down to the \textit{Swift} detection flux threshold of 1.5 ph s$^{-1}$ cm$^{-2}$.  Since the lowest-redshift \textit{Swift}-detected source is at $z=0.1$, the peak flux function is not well constrained below a source flux of 1.5 ph s$^{-1}$ cm$^{-2}$ at $z=0.1$, i.e., an integrated isotropic source flux of $3.5\times 10^{55}$ ph s$^{-1}$.  \textbf{Right} panel: predicted apparent cumulative peak flux functions for sGRBs from the flux distribution in the left panel and the redshift distribution implied by three different time delay models.  Compared to data from the most recent Fermi and BATSE catalogs, these apparent peak flux functions are in excellent agreement; Kolmogorov-Smirnov tests show no discrepancies ($p>0.9$).  All units are normalized to fluxes in the BATSE energy band (50--300 keV).}
\label{f:grb_lf}
\end{figure}

\section{Determining the Redshift Sensitivity of \textit{Swift}}
\label{a:sensitivity}

The observed redshift distribution of sGRBs from \textit{Swift} \citep{Fong13} and the apparent luminosity or peak flux function \citep{Goldstein12,Goldstein13} may be combined to give a constraint on the intrinsic sGRB flux function \citep{Guetta06}.  Equivalently, they give a constraint on the redshift sensitivity of \textit{Swift} \citep{Guetta06,Coward12,Kelley13}, under the assumption that the intrinsic flux distribution of sGRBs is independent of redshift.  With this assumption, the intrinsic cumulative flux function (non beaming-corrected) is separable into a flux-dependent shape and a redshift-dependent number density:
\begin{equation}
\Phi_\mathrm{int}(F,z) = N_\mathrm{int}(z) \Phi_\mathrm{int}(F)
\end{equation}
The observed redshift distribution, $P_\mathrm{obs}(z)$, and the apparent cumulative flux function, $\Phi_\mathrm{obs}(F)$ can then be written as
\begin{eqnarray}
P_\mathrm{obs}(z) & = & \frac{V'(z)}{1+z}N_\mathrm{int}(z) \Phi_\mathrm{int}(F_\mathrm{min}\mu_D(z)k(z)) \label{e:pz}\\
\Phi_\mathrm{obs}(F) & = & \int_0^\infty \frac{V'(z)}{1+z}N_\mathrm{int}(z) \Phi_\mathrm{int}(F\mu_D(z)k(z))dz \label{e:phi_obs}
\end{eqnarray}
where $V(z)$ is the comoving volume out to redshift $z$, $L_\mathrm{min}$ is the minimum detectable absolute luminosity, $\mu_D(z)$ is the cosmological dimming factor, and $k(z)$ is the $k$-correction.\footnote{Note that the relevant cosmological dimming factor in this case is $\mu_D(z) = D_L(z)^2 (1+z)^{-1}$, where $D_L(z)$ is the luminosity distance; the extra factor of $(1+z)^{-1}$ is because the detector threshold is based on photon \textit{counts} instead of photon \textit{energies}.}  Finally, the redshift sensitivity, $\epsilon_\mathrm{detector}(z)$, is
\begin{equation}
\label{e:detector_fit}
\epsilon_\mathrm{detector}(z) = \frac{\Phi_\mathrm{int}(F_\mathrm{min}\mu_D(z)k(z))}{\Phi_\mathrm{int}(F_\mathrm{min})}.
\end{equation}
Eqs.\ \ref{e:pz} and \ref{e:phi_obs} are typically solved for $\Phi_\mathrm{int}(F)$ and $N_\mathrm{int}(z)$ via forward deconvolution \citep{Guetta06}.  However, since \cite{Guetta06} have already found that a wide range of power-law time-delay models give acceptable fits, we can take a simpler approach.  We first assume that $N_\mathrm{int}(z)$ is given by the $P(\Delta t) \propto (\Delta t)^{-1}$ time-delay model; this allows $\epsilon_\mathrm{detector}(z)$ to be determined directly from the observed sGRB redshift distribution (Eq.\ \ref{e:pz}).  Using this assumed $\epsilon_\mathrm{detector}(z)$, we then check that the implied apparent flux function (Eq.\ \ref{e:phi_obs}) is consistent with observations for all three of the time-delay models.

The first step requires calculating the $k$-correction for sGRBs, as well as the knowing the minimum observable flux $F_\mathrm{min}$ for \textit{Swift}.  Following \cite{Coward12}, we adopt a \cite{Band93} spectral fitting function for sGRBs, with parameters $\alpha = -1$, $\beta = -2.3$, and $E_0 = 511$ keV.  In addition, we adopt the same effective minimum flux limit of 1.5 ph s$^{-1}$ cm$^{-2}$ for \textit{Swift} sGRBs.  We find that the resulting \textit{Swift} redshift sensitivity is extremely well-fit by
\begin{equation}
\label{e:detector_fit_final}
\epsilon_\mathrm{detector}(z) = \exp(-4.3z)
\end{equation}
assuming that the \cite{Fong13} redshift distribution is unbiased (see discussion in Appendix \ref{a:redshift_incomp}).  This functional form, as well as the comparison to the observed redshift distribution from \cite{Fong13} are both shown in Fig.\ \ref{f:grb_weighted}.  The fit for the modeled redshift distribution for the $P(\Delta t) \propto (\Delta t)^{-1}$ time-delay model is extremely good; a K-S test finds negligible discrepancy ($p>0.9$).  Using this same fit for the $P(\Delta t) \propto (\Delta t)^{-0.5}$ and $(\Delta t)^{-1.5}$ is still more than acceptable (Fig.\ \ref{f:grb_weighted}); K-S tests for both give $p$-values greater than 0.48.

Combining Eqs.\ \ref{e:detector_fit} and \ref{e:detector_fit_final}, we find an implicit equation for the intrinsic cumulative flux function:
\begin{equation}
\label{e:intrinsic_lf}
\Phi_\mathrm{int}(F\mu_D(z)k(z)) = \Phi_\mathrm{int}(L_\mathrm{min}) \exp(-4.3z)
\end{equation}
The resulting functional form for $\Phi_\mathrm{int}(F)$ is shown in Fig.\ \ref{f:grb_lf}.  Since the minimum \textit{Swift}-discovered sGRB redshift is $z\approx 0.1$ \citep{Fong13}, the intrinsic flux function is not well-constrained for apparent source fluxes below 1.5 ph s$^{-1}$ cm$^{-2}$ for a source at $z=0.1$.

From Eqs.\ \ref{e:intrinsic_lf} and \ref{e:phi_obs}, as well as the redshift distributions implied by the three time-delay models, we have calculated the expected apparent flux functions (Fig.\ \ref{f:grb_lf}, right panel).  For comparison, we have also plotted the actual apparent flux functions obtained by combining the BATSE 5B \citep{Goldstein13} and Fermi two-year catalogs \citep{Goldstein12} for bursts with $T_{90} < 2$s.  Since the BATSE 5B spectral catalog did not report 64ms peak fluxes, we multiplied the 2s fluxes by $(2$s$ / T_{90})$ to obtain the effective peak flux.  In addition, we used the Fermi 64ms peak fluxes matched to the BATSE spectral energy window.

As can be seen from Fig.\ \ref{f:grb_lf}, the agreement between the modeled and observed sGRB flux functions is remarkable.  Regardless of the time-delay model, K-S tests show no discrepancies ($p>0.9$ in all cases).  Hence, we adopt the fitting formula in Eq.\ \ref{e:detector_fit_final} as the effective \textit{Swift} detector sensitivity.

\begin{figure}
\plotminigrace{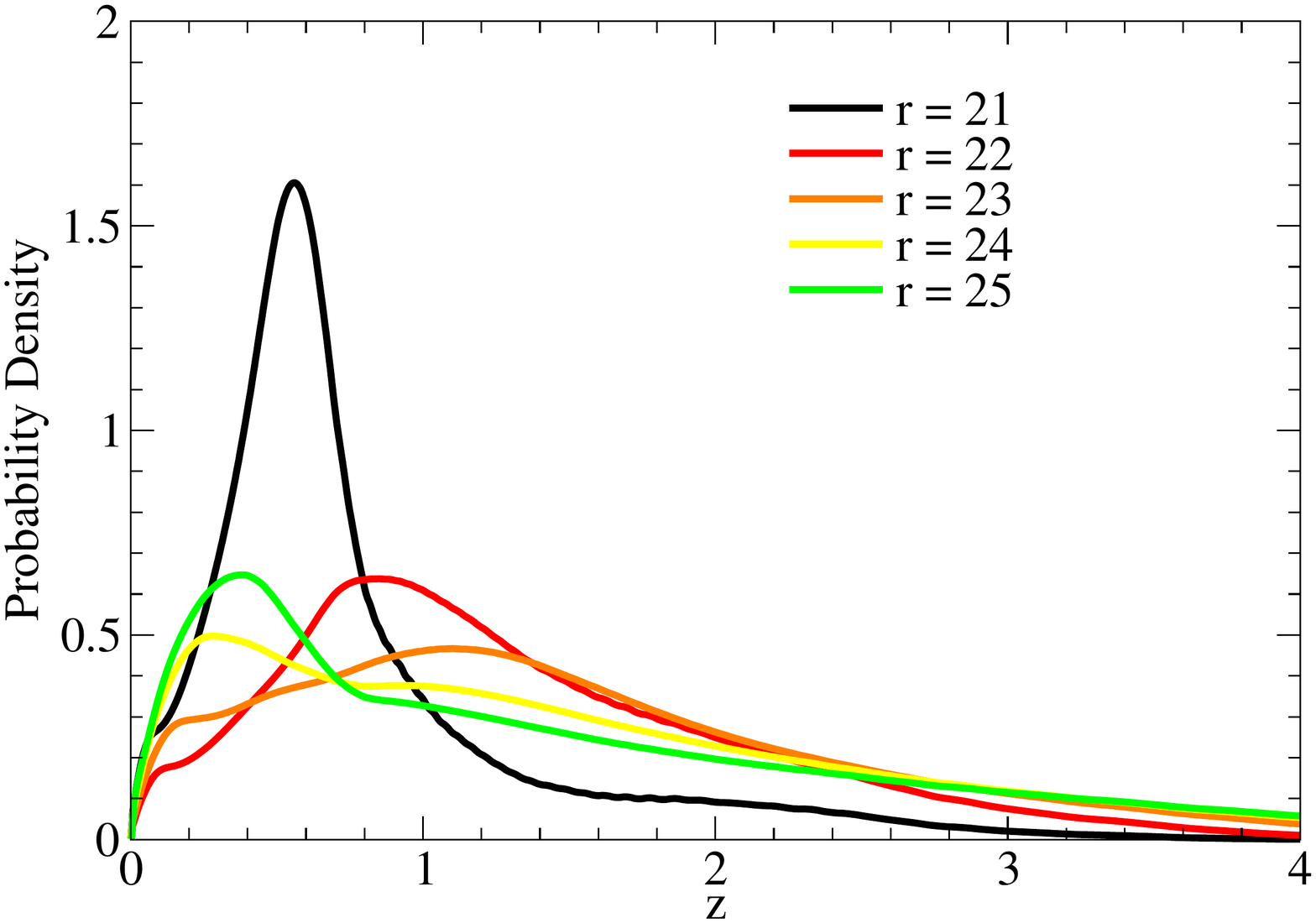}\plotminigrace{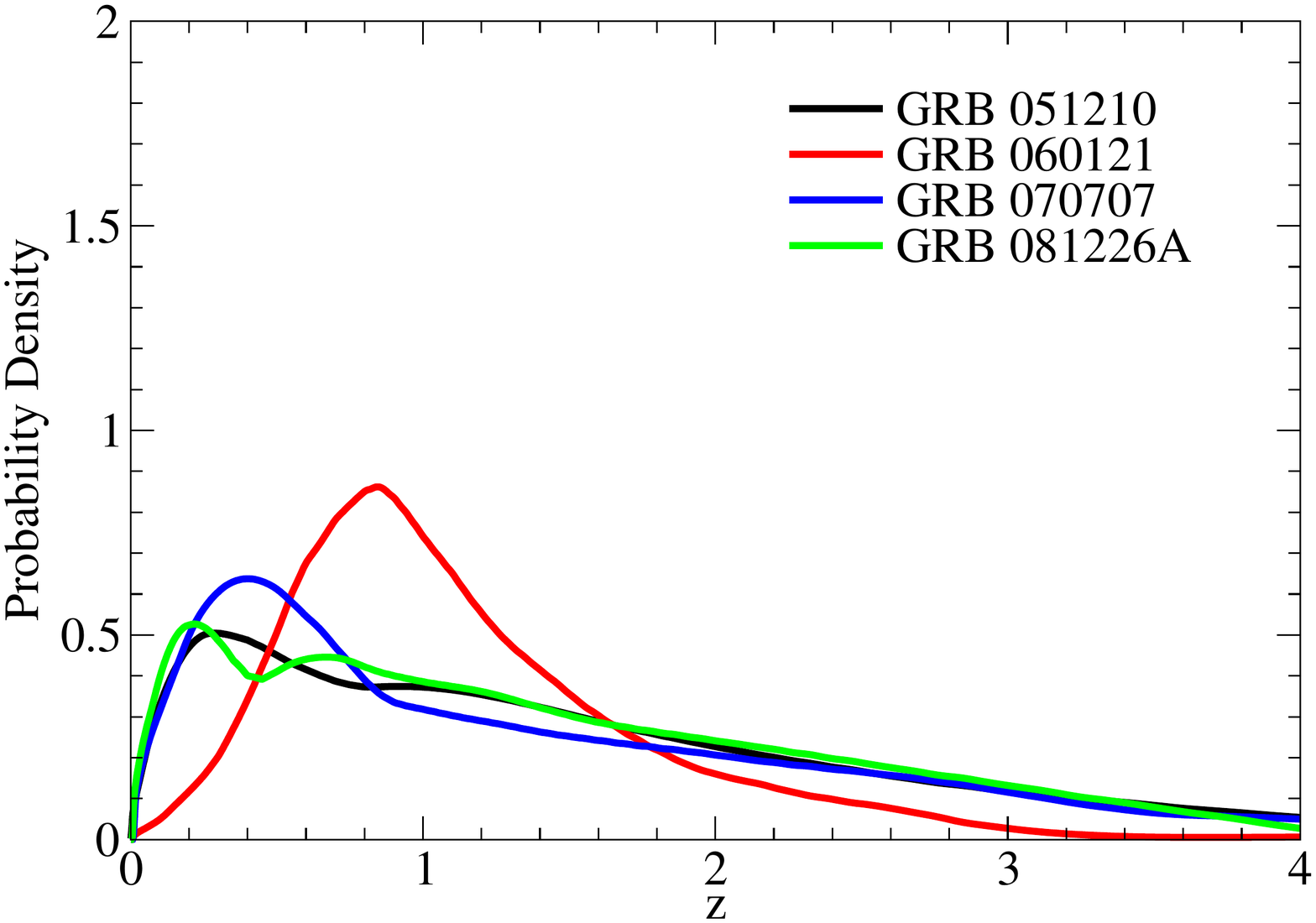}
\caption{\textbf{Left} panel: Redshift probability distributions for galaxies as a function of their SDSS $r$-band apparent magnitudes (AB). A fixed apparent magnitude corresponds to a brighter absolute magnitude, at which there will be fewer galaxies, so the redshift distribution is tightly correlated with the shape of the luminosity function; fainter galaxies are not necessarily at higher redshifts (see text).  \textbf{Right} panel: probability distributions for the redshifts of the host galaxies given available magnitudes for GRB 051210 ($r=24.04$), 060121 (F606W$=27.0$, F160W$=24.5$), 070707 (F606W$=26.16$, F160W$=26.16$), and 081226A ($r'=25.79$, $g'=25.88$).}
\label{f:z_dist}
\end{figure}

\section{Redshift Incompleteness}

\label{a:redshift_incomp}

The observed redshift distribution used in Fig.\ \ref{f:grb_weighted} is derived from associated host galaxy redshifts for 26 sGRBs with accurate positions \citep{Fong13}.  For an additional 4 sGRBs, the associated host galaxies do not have available redshifts \citep[051210, 060121, 070707, and 081226A;][]{Berger07,Levan06,Piranomonte08,Guelbenzu12}.  In one case (051210), available spectra are featureless, suggesting that this sGRB may lie in the redshift desert ($1.4 < z < 2.5$).  In the three other cases, the host galaxies were too faint for straightforward spectral followup.

We have investigated what redshift information may be extracted from available measurements of these host galaxies.  Knowing the apparent magnitude of a galaxy provides a weak prior on its redshift (Fig.\ \ref{f:z_dist}), determinable from the redshift dependence of the galaxy luminosity function and the differential observable volume.  We note that, while high-redshift galaxies are universally faint, faint galaxies are not all at high redshifts.  Because of the upturn in the faint-end slope of the luminosity function \citep{Blanton05}, very faint galaxies ($r>24$) are more likely to be at low redshifts than merely faint galaxies ($22<r<24$).  Using galaxy luminosities derived in the same way as discussed in \S \ref{s:biases}, we have calculated the resulting posterior redshift distributions given existing luminosity measurements for the hosts of the four sGRBs with missing redshifts (Fig.\ \ref{f:z_dist}, right panel).

Encouragingly, all these distributions peak in the same redshift window as occupied by the better-measured sGRB sample (Fig.\ \ref{f:grb_weighted}).  On the other hand, the distributions are very broad.  If in fact all of these galaxies occurred at high redshifts (e.g., $z>1.5$), the main impact would be to boost the inferred \textit{Swift} sensitivity at $z\sim 2$ by a factor of 5, to 0.1\%.  Correspondingly, the intrinsic luminosity function inferred in Appendix \ref{a:sensitivity} would be boosted at the luminous end.  However, we note that the ``Upgraded \textit{Swift}'' model considered in Appendix \ref{s:quant} tests the effect of boosting $z=2$ sensitivity by a factor of 74, with only a minimal effect on our results in the main body of the paper.

\end{document}